\documentclass[aps,pra,twocolumn,export]{revtex4-1} 
\usepackage[colorlinks=true,linkcolor=blue,citecolor=blue,urlcolor=blue]{hyperref}

\usepackage[utf8]{inputenc}
\usepackage[german, english]{babel}

\usepackage{dcolumn}  
\usepackage{bm}        
\usepackage{bbold}
\usepackage{subfigure}
\usepackage{amsthm}
\usepackage{mathtools}
\usepackage{tikz}
\usetikzlibrary{shapes.geometric, arrows, shapes.multipart}
\usepackage{xcolor}
\usepackage{booktabs, tabularx}

\usepackage[export]{adjustbox}
\usepackage{titlesec}
\usepackage{amsfonts}
\usepackage{amssymb}
\usepackage{braket}

\begin{document}

\title{Subsystem symmetry enriched topological order in three dimensions}
\author{David T. Stephen}
\affiliation{Max-Planck-Institut f{\"u}r Quantenoptik, Hans-Kopfermann-Stra{\ss}e 1, 85748 Garching, Germany}
\affiliation{Munich Center for Quantum Science and Technology, Schellingstra{\ss}e 4, 80799 M{\"u}nchen, Germany}
\author{ Jos\'e Garre-Rubio}
\affiliation{Departamento de An\'alisis Matem\'atico  y Matem\'atica Aplicada, UCM, 28040 Madrid, Spain} 
\affiliation{ICMAT, C/ Nicol\'as Cabrera, Campus de Cantoblanco, 28049 Madrid, Spain}

\author{Arpit Dua}
  \affiliation{Department of Physics, Yale University, New Haven, Connecticut 06511, USA}
  \affiliation{Yale Quantum Institute, Yale University, New Haven, Connecticut 06511, USA}

\author{Dominic~J. Williamson}
\affiliation{Stanford Institute for Theoretical Physics, Stanford University, Stanford, CA 94305, USA}

\date{\today}

\begin{abstract}
We introduce a model of three-dimensional (3D) topological order enriched by planar subsystem symmetries. The model is constructed starting from the 3D toric code, whose ground state can be viewed as an equal-weight superposition of two-dimensional (2D) membrane coverings. We then decorate those membranes with 2D cluster states possessing symmetry-protected topological order under line-like subsystem symmetries. This endows the decorated model with planar subsystem symmetries under which the loop-like excitations of the toric code fractionalize, resulting in an extensive degeneracy per unit length of the excitation. We also show that the value of the topological entanglement entropy is larger than that of the toric code for certain bipartitions due to the subsystem symmetry enrichment. Our model can be obtained by gauging the global symmetry of a short-range entangled model which has symmetry-protected topological order coming from an interplay of global and subsystem symmetries. We study the non-trivial action of the symmetries on boundary of this model, uncovering a mixed boundary anomaly between global and subsystem symmetries. To further study this interplay, we consider gauging several different subgroups of the total symmetry. The resulting network of models, which includes models with fracton topological order, showcases more of the possible types of subsystem symmetry enrichment that can occur in 3D.
\end{abstract}

\maketitle

\section{Introduction}

A new paradigm in the classification of gapped phases of matter has recently begun thanks to the discovery of models with novel sub-dimensional physics. This includes the fracton topological phases \cite{Chamon2005,Haah2011,doi:10.1080/14786435.2011.609152,kim20123d,yoshida2013exotic,Vijay2015,Vijay2016,Williamson2016,Shirley2018,
Prem2019a,Nandkishore2019,Pretko2020}, in which topological quasi-particles are either immobile or confined to move only within subsystem such as lines or planes, as well as models with subsystem symmetries, which are symmetries that act non-trivially only on rigid subsystems of the entire system \cite{Nussinov2009,Nussinov2009a,Vijay2016,Williamson2016,Shirley2019,You2018,You2018a,Devakul2018,Devakul2019,Schmitz2019, Devakul2019a,Stephen2019,Doherty2009,Else2012a,Raussendorf2003,Raussendorf2019,
Devakul2018a,Stephen2019a,Daniel2019}. These two types of models are dual since gauging subsystem symmetries can result in fracton topological order, analogous to how topological order can be obtained by gauging a global symmetry \cite{Vijay2016,Williamson2016,Shirley2019}. Models with subsystem symmetries are also interesting in their own right. For example, one may use them to define symmetry-protected topological (SPT) phases, giving rise to the so-called subsystem SPT (SSPT) phases \cite{Raussendorf2019,You2018,You2018a,Devakul2018,Devakul2019,Schmitz2019,
Devakul2019a,Stephen2019a,Stephen2019}. SSPT phases display new physics compared to conventional SPT phases such as extensive edge degeneracy and unique entanglement properties. They also serve as resources for quantum computation via the paradigm of measurement-based quantum computation \cite{Doherty2009,Else2012a,Raussendorf2003,Raussendorf2019,
Devakul2018a,Stephen2019a,Daniel2019}.

In this paper, we build on this paradigm of sub-dimensional physics by considering how topological order may be \textit{enriched} in the presence of subsystem symmetries. We call such order \textit{subsystem symmetry-enriched topological} (SSET) order.
For global symmetries, the theory of symmetry-enriched topological (SET) phases of matter is well developed, and is characterized by a non-trivial action of the symmetry on the topological excitations~\cite{Mesaros2013,Barkeshli2013a,teo2015theory,tarantino2015symmetry,Barkeshli2019,barkeshli2016reflection}. In 2D, these excitations are point-like anyons. The symmetry can act on the anyons by permuting them, and the anyons may also carry a fractional symmetry charge, which means that the action of the symmetry on a state containing several anyons reduces to non-trivial projective actions localized around each anyon. For example, in spin liquids with $SO(3)$ spin rotation symmetry, the whole system forms a spin singlet, but individual anyons, \textit{i.e.} spinons, may have non-trivial spin, and hence carry a degeneracy that is protected by the symmetry. Crucially, single anyons cannot be created, and any physical state containing some anyons will transform under symmetry overall in a trivial manner, despite the non-trivial action on each anyon. In 3D, the situation is made more complicated by the presence of extended loop-like excitations, but significant progress has nonetheless been made \cite{Cenke2013,Cheng2015,Chen2016,Ning2016,Fidkowski2017,Lan2018,Lan2019}.

In the presence of subsystem symmetries, the kind of fractionalization on mobile point-like excitations described in the preceding paragraph cannot occur. In anticipation of the main model of this paper, consider a 3D system with planar subsystem symmetries, and suppose there are two point-like excitations living on a single plane. It may seem as though it is possible for each excitation to carry a fractional charge, as above. However, in this case, we can simple move one of the excitations away from this plane, such that the subsystem symmetry now only acts on a single excitation. Since the total action on the system must be trivial, the local action on a single excitation must also be trivial, and there is no fractionalization or permutation. One way around this is to consider excitations with restricted mobility, as in fracton topological order. In this case, it may not be possible to move an excitation away from the symmetry plane, or there may be a conservation law that restricts the number of excitations on each plane to be \textit{e.g.} even, so fractionalization again becomes possible. Such an option was explored in Ref.~\cite{You2018a}. 

\begin{figure}[t]
\centering
\subfigure[]{\raisebox{3mm}{\label{fig:leada}\includegraphics[scale=0.1]{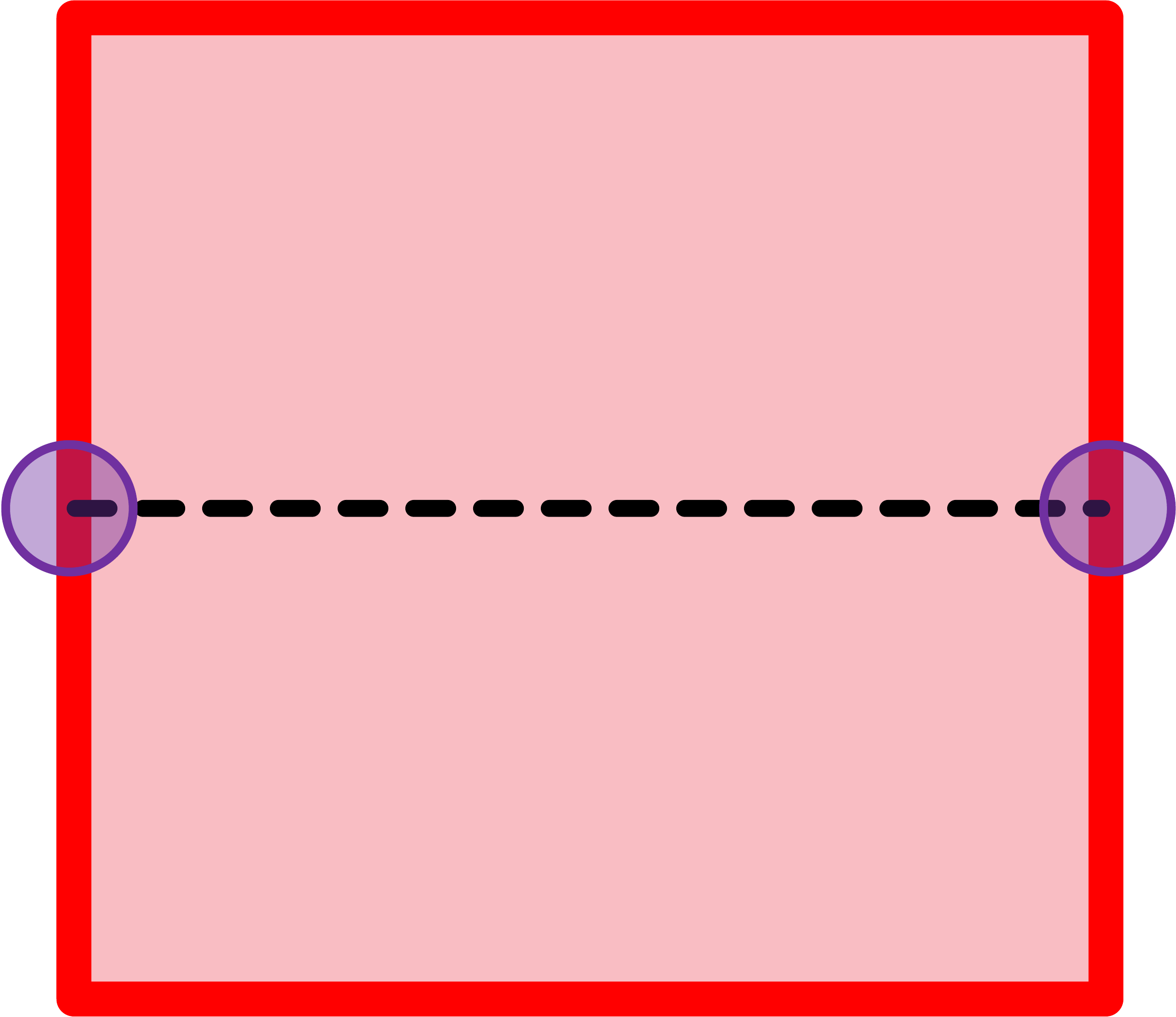}}} \hfill
\subfigure[]{\label{fig:leadb}\includegraphics[scale=0.1]{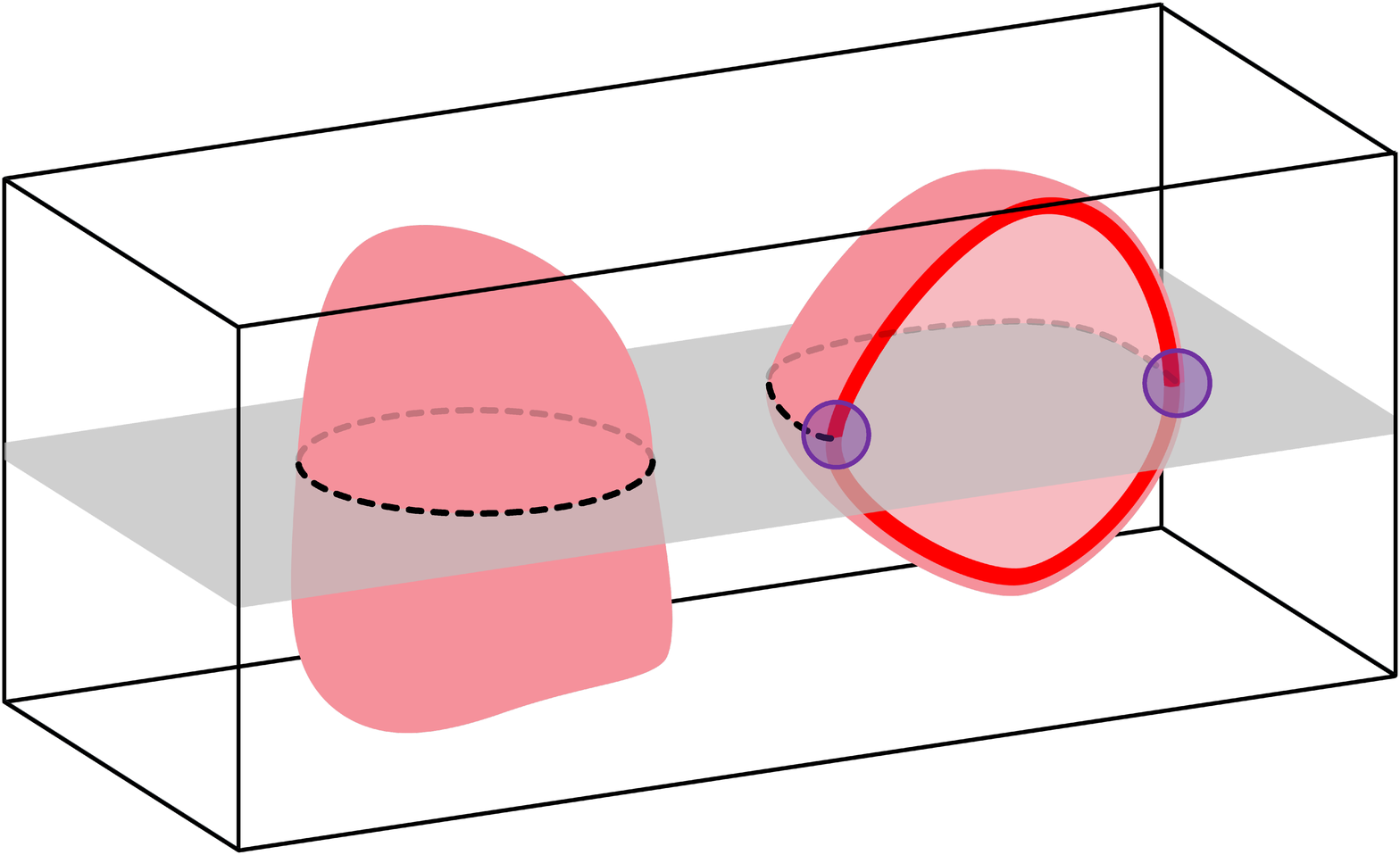}}
\caption{(a) Schematic diagram of a 2D SSPT order with line-like symmetries (dashed line) in the presence of a boundary (thick red line). Near a boundary, neighbouring line symmetries locally anti-commute (small circles), leading to an extensive degeneracy on the edge. (b) The ground state of the 3D SSET order has planar subsystem symmetries, and is a condensate of closed membranes of 2D SSPT orders. Left: a closed membrane of 2D SSPT order. Planar operators intersect the membrane along a loop, indicated by the dashed line, and reduce to the line symmetries of the membrane. Right: an open membrane carries a line-like excitation on its boundary, indicated by the thick red line, which transforms non-trivially under subsystem symmetries in the same way as the boundary of the 2D SSPT order.}
\label{fig:lead}
\end{figure}

In this paper, we show that topological excitations of higher spatial dimension lead to another opportunity for subsystem symmetry fractionalization. Namely, consider a 1D loop-like excitation in a 3D system. A global symmetry cannot fractionalize on the loop, at least not in the simple way described above, since there is no way to decompose the symmetry action as a product of disjoint local actions. However, when we consider subsystem symmetries, fractionalization on a single loop is possible. Consider a plane which intersects a loop-like excitation. Acting with a symmetry on the plane reduces to a product of local actions at each intersection point. Since the number of intersection points is necessarily even, it is possible for the symmetry to fractionalize at each point. This is precisely what occurs in the model described in this paper. Additionally, we find models containing both point-like excitations with restricted mobility, as well as loop-like excitations, and a non-trivial symmetry action that couples the two types.

\subsection{Summary of results}

Our SSET model can be understood as a decorated 3D toric code model. \cite{Hamma2005a,Castelnovo2008}. This toric code consists of qubits on the faces of a cubic lattice, and the ground states can be visualized as equal-weight superpositions over all basis states where the faces in the state $|1\rangle$ form unions of closed 2D membranes. To enrich this model with subsystem symmetries, we introduce new qubits on the edges of the lattice and couple them to the faces in such a way that edge qubits lying on membranes form 2D cluster states, and those away from membranes remain in symmetric product states. The 2D cluster state \cite{Raussendorf2003} has line-like subsystem symmetries and is the prototypical example of SSPT order \cite{Raussendorf2019,You2018}, see Fig.~\ref{fig:leada}. Constructed in this way, the decorated model has planar subsystem symmetries acting on the decorating qubits. This is due to the fact that, along the intersections of a given plane and the membranes, the symmetry action of this plane reduces to the line-like symmetries of the decorating cluster states, see Fig~\ref{fig:leadb}. Furthermore, the loop excitations of the toric code, which appear along the boundary of open membranes, now coincide with the boundary of cluster states. Due to the SSPT order of cluster states, these boundaries, and hence the loop excitations, fractionalize under the planar symmetries. The immediate consequence of this fractionalization is that the loop excitations are endowed with a degeneracy per unit length which is protected by the subsystem symmetry. The procedure of decorating topological orders with lower-dimensional SPT orders is a well-established way to create SET orders \cite{Chen2014,Huang2014,Li2014,Ben-Zion2016}, and our construction here can be seen as a generalization of this procedure to subsystem symmetries, with the notable feature that $d$-dimensional subsystem symmetries of the decorating SSPT translate into $(d+1)$-dimensional subsystem symmetries of the decorated model.

Another effect of the subsystem symmetry fractionalization that we discover is an increased value of the topological entanglement entropy (TEE). The TEE refers to a constant correction to the area law when computing bipartite entanglement entropy, and is a topological invariant, in the sense that it takes a uniform value within a given topological phase of matter \cite{Kitaev2006,Levin2006}. Recently, it has been understood that there are certain quantum states for which the TEE, for certain bipartitions, does not match the expected value  \cite{Cano2015,Zou2016,Santos2018,Devakul2018,Williamson2019,Schmitz2019,Stephen2019,Kato2019}. In a majority of known cases (see Ref.~\cite{Kato2019} for a possible counterexample), this is due to the presence of lower dimensional SPT order around the boundary of the bipartition, which can in turn be related to the presence of SSPT order \cite{Devakul2018}. In fact, it was shown in Ref.~\cite{Stephen2019} that this ``spurious'' TEE takes a uniform value within the SSPT phase of the 2D cluster state, and can be used to detect SSPT order and phase transitions. To emphasize its relation to the SSPT order, it was dubbed the symmetry-protected entanglement entropy (SPEE). For the case of an SSET, which naturally has a non-zero TEE coming from the topological order, one might therefore expect a larger value of the TEE due to the additional presence of the SPEE. This is exactly what we confirm in our SSET model.

\begin{figure}
\centering
\tikzstyle{block1} = [rectangle, fill = gray!25, minimum width=2.5cm, minimum height=0.7cm, text centered, thick, draw=black, rounded corners]
\tikzstyle{block2} = [rectangle, fill = gray!25, minimum width=2.5cm, minimum height=0.7cm, text centered, thick, draw=black, rounded corners]

\tikzstyle{arrow} = [thick,->,>=stealth, shorten <=3pt,shorten >=3pt]
\begin{tikzpicture}[node distance=2cm,every text node part/.style={align=center}]

\node (sspt) [block1] {\textbf{SSPT} \\ Sec.~\ref{sec:sspt}};
\node (sset) [block1, below of=sspt, xshift = 0 cm] {\textbf{SSET} \\ Sec.~\ref{sec:sset}};
\node (frac1) [block1, left of=sset, xshift = -1.0 cm] {\textbf{SSE Fracton} \\ Sec.~\ref{sec:frac1}};
\node (frac2) [block1, right of=sset, xshift = 1.0 cm] {\textbf{SSE Fracton} \\ Sec.~\ref{sec:frac2}};
\node (frac4) [block1, below of=sset, xshift = 0 cm] {\textbf{SE Fracton} \\ Sec.~\ref{sec:frac3}};
\node (frac5) [block1, right of=frac4, xshift = +1.0 cm] {\textbf{SSE Panoptic} \\ Sec.~\ref{sec:pan2}};
\node (frac3) [block1, left of=frac4, xshift = -1.0 cm] {\textbf{SSE Panoptic} \\ Sec.~\ref{sec:pan1}};
\node (pan)  [block1, below of=frac4, xshift = 0 cm] {\textbf{Panoptic} \\ Sec.~\ref{sec:pan3}};
\draw [arrow] (sspt) -- (sset) node[midway, anchor=east] {$g$};
\draw [arrow] (sspt) -- (frac2) node[midway,sloped,above] {$s_2$};
\draw [arrow] (sspt) -- (frac1) node[midway,sloped,above] {$s_1$};
\draw [arrow] (frac1) -- (frac3) node[midway,anchor=east] {$g$};
\draw [arrow] (frac2) -- (frac5) node[midway,anchor=west] {$g$};
\draw [arrow] (sset) -- (frac3) node[pos=0.75,sloped,above] {$s_1$};
\draw [arrow] (sset) -- (frac5) node[pos=0.75, sloped, above] {$s_2$};
\draw [arrow,preaction={draw, line width=3pt, white}] (frac1) -- (frac4) node[pos=0.75,sloped,above] {$s_2$};
\draw [arrow,preaction={draw, line width=3pt, white}] (frac2) -- (frac4) node[pos=0.75,sloped,above] {$s_1$};
\draw [arrow] (frac3) -- (pan) node[midway,sloped,above] {$s_2$};
\draw [arrow] (frac4) -- (pan) node[midway,anchor=east] {$g$};
\draw [arrow] (frac5) -- (pan) node[midway,sloped,above] {$s_1$};

\end{tikzpicture}
\caption{Flowchart describing the various models obtained by gauging and ungauging the symmetries of the SSET in different ways. The arrows are labelled by the symmetry which is gauged along the direction they point, where $g$, $s_1$, and $s_2$ represent the global symmetry, lattice-plane subsystem symmetries, and dual-plane subsystem symmetries, respectively. A model is called ``fracton'' if all excitations display some mobility restrictions (this includes stacks of 2D topological orders), and ``panoptic'' if restricted-mobility excitations appear alongside fully mobile and loop-like excitations. ``SE'' denotes models with symmetry enrichment due to global symmetries alone, while ``SSE'' denotes models enriched by subsystem symmetries, possibly alongside global symmetries.}
\label{fig:flow}
\end{figure}
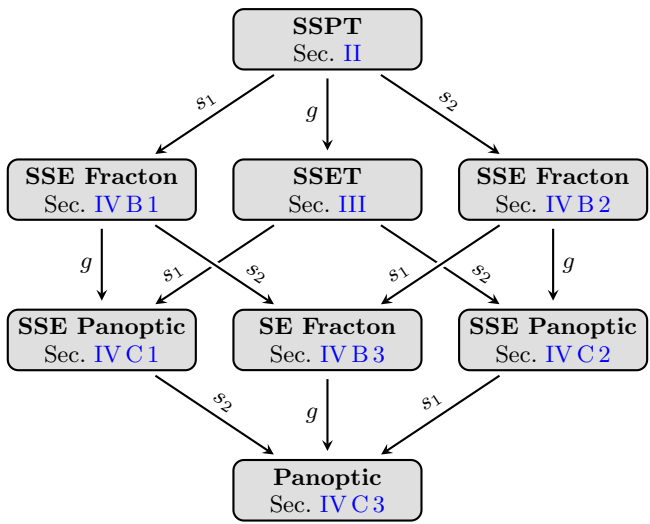

Our SSET model stems from a unique interplay of global and subsystem symmetries. To better understand this interplay, we consider gauging and ungauging the symmetries of the SSET in various combinations, resulting in a network of eight different models, as pictured in Fig.~\ref{fig:flow}. At the root of this network is a short range entangled model with SSPT order under a combination of global and subsystem symmetries. We calculate the cocycle that encodes the non-trivial action of the symmetries on the boundary of this model, revealing a mixed anomaly between global and subsystem symmetries. Using this cocycle, we calculate the effect of the symmetries on the extrinsic symmetry defects, which allows us to predict the outcome of gauging the symmetries in different combinations. We then discuss the nature of the topological order and the symmetry enrichment of each of the eight models, and in particular show that the model resulting from gauging all symmetries, \textit{i.e.} the model resulting from gauging the subsystem symmetries of the SSET, displays the ``panoptic'' order that was recently discovered in Refs.~\cite{Prem2019,Bulmash2019}, and contains non-abelian fractons.

The rest of the paper is structured as follows. In Section \ref{sec:sspt}, we introduce a model with both global and subsystem symmetries, and show that it has non-trivial SPT order by analysing its boundary. Then, in Section \ref{sec:sset}, we gauge the global symmetries of this model to obtain our model of SSET order and study its properties. In Section \ref{sec:gauging}, we consider gauging the subsystem symmetries and investigate the network of models in Fig.~\ref{fig:flow}. Finally, in Section \ref{sec:conclusion}, we discuss some principles that could help guide a general theory of subsystem symmetry enrichment, give an argument against the existence of SSET order in 2D, and discuss possible routes of future work.

\section{Symmetry protected topological order with global and subsystem symmetries} \label{sec:sspt}

To derive our model of SSET order, we begin with a short-ranged entangled model that has SPT order with respect to a combination of global and subsystem symmetries. The SSET is obtained by gauging only the global symmetries of this model. The process of partially gauging symmetries of an SPT order to obtain SET order is well understood in the case of global symmetries \cite{Barkeshli2019,Garre-Rubio2017,NewSETPaper2017,Lan2019}, and is reviewed in a simple 2D example in Appendix~\ref{app:2d}. 

We first define the short-range entangled model, and demonstrate its non-trivial SPT order by identifying the non-trivial action of the symmetries on the boundary, as encoded by a certain 3-cocycle. Notably, the non-trivial order arises from an interplay between the global and subsystem symmetries, and the system becomes trivial if only the global symmetry or only the subsystem subsystem symmetry is preserved. We will refer to this type of order as SSPT order, despite the equal importance of both global and subsystem symmetries.

\begin{figure}[t]
\centering
\subfigure[]{\label{fig:ssptham}\includegraphics[scale=0.087]{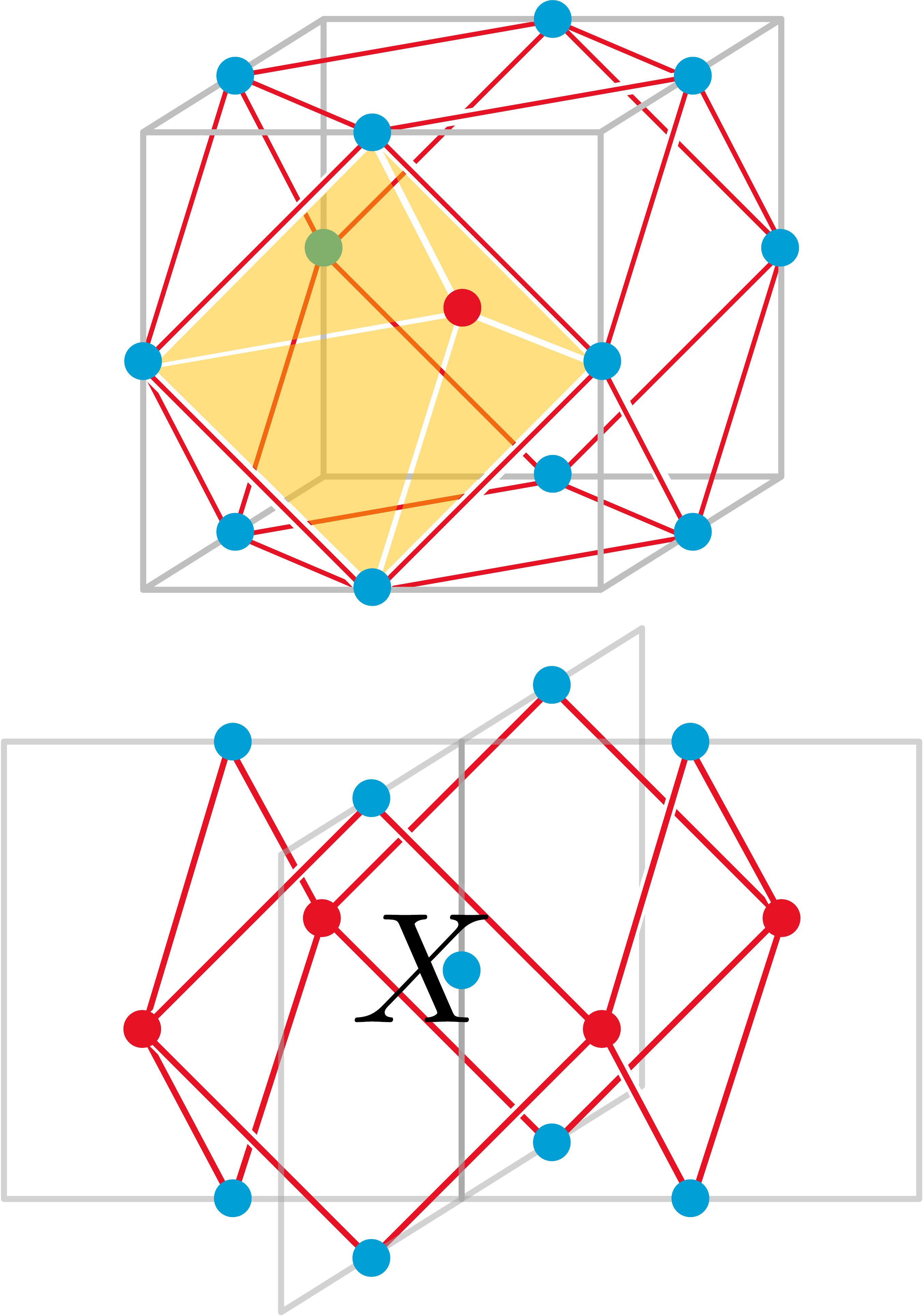}} \hfill
\subfigure[]{\label{fig:clustersoup}\includegraphics[scale=0.087]{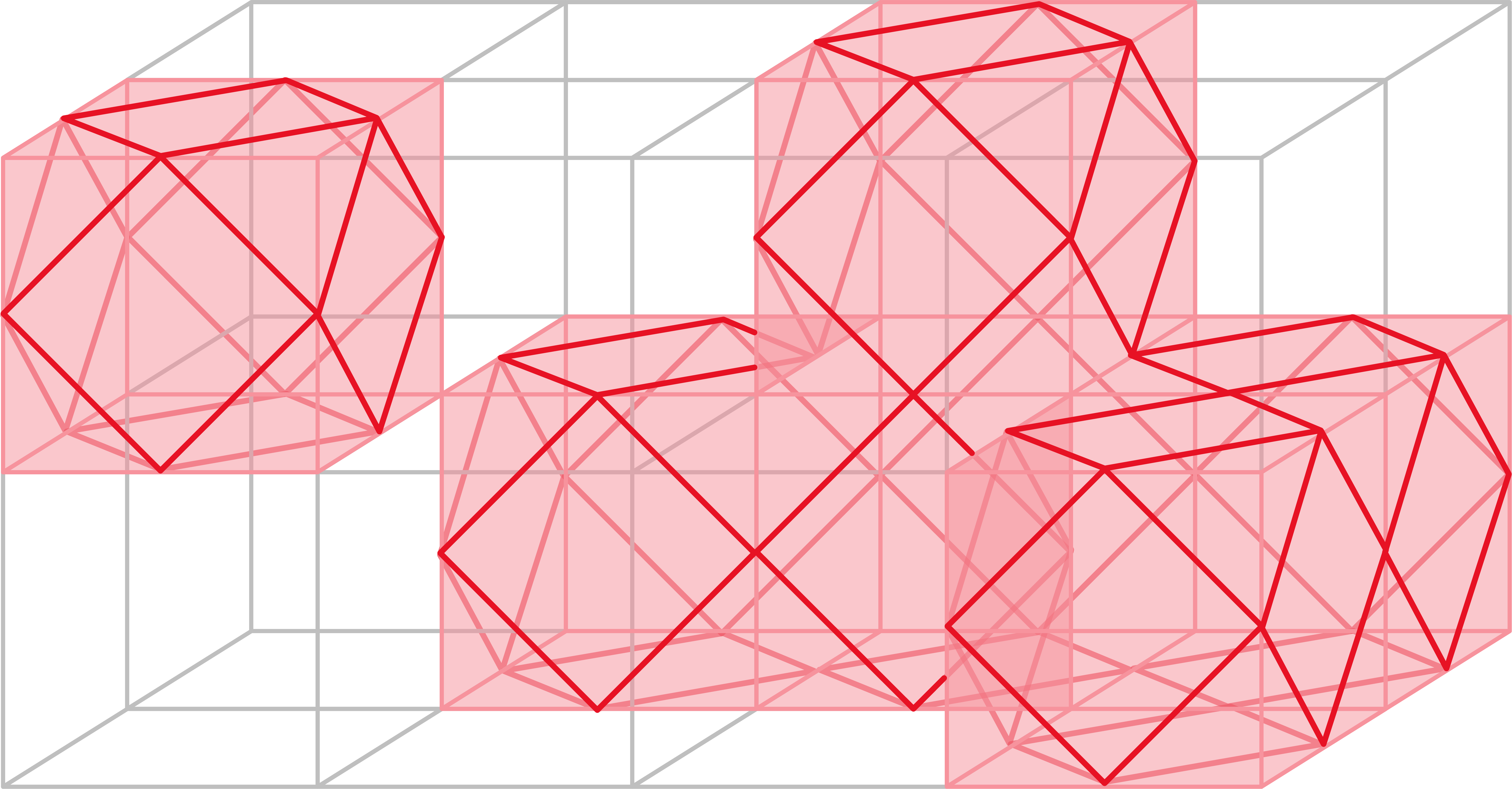}} 
\caption{(a) Top: a unit cell of the 3D SSPT state. Four of the triangles $\triangle$ appearing in Eq.~(\ref{eq:uccz}) are shown; there are four such triangles for each face within this cell. The red lines represent the $CZ$ gates contained in $U_c$ (Eq.~(\ref{eq:uc})). Bottom: the Hamiltonian term $\widetilde{C}_e$, where $e$ is the central vertical edge. (b) Decoration by 2D cluster states. The colored faces depict either domain walls, in the case of $|SSPT\rangle$, or membranes, in the case of $|SSET\rangle$.  }
\end{figure}

The SSPT model lives on a simple 3D cubic lattice, with qubits in the body centers ($C$) and on the edges ($E$). To begin, consider a trivial paramagnetic Hamiltonian acting on this system,
\begin{equation}
H_{triv}=-\sum_{c\in C} X_c -\sum_{e\in E} X_e\ ,
\end{equation}
whose unique ground state is a product state of $|+\rangle = \frac{1}{\sqrt{2}}(|0\rangle + |1\rangle)$ on every edge and body qubit.
Our model can be defined by acting on this trivial system with a finite depth unitary circuit,
\begin{equation} \label{eq:uccz}
U_{CCZ}=\prod_{\triangle} CCZ_\triangle
\end{equation}
where the product runs over all triples of qubits $\triangle$ consisting of one body qubit and two of its nearest-neighbouring edge qubits, as pictured in Fig.\ref{fig:ssptham}, and $CCZ_\triangle$ acts on the three qubits as $CCZ|i\rangle |j\rangle |k\rangle=(-1)^{ijk}|i\rangle |j\rangle |k\rangle$. Using the fact that $CCZ_{abc} X_a CCZ_{abc}^\dagger = X_a CZ_{bc}$, where $CZ|i\rangle |j\rangle=(-1)^{ij}|i\rangle |j\rangle$, we obtain the Hamiltonian,
\begin{align}
H_{SSPT}&=U_{CCZ} H_{triv} U_{CCZ}^\dagger \nonumber \\
&=-\sum_{c\in C} \widetilde{B}_c -\sum_{e\in E} \widetilde{C}_e,
\end{align}
where $\widetilde{C}_e \equiv U_{CCZ} X_e U_{CCZ}^\dagger$ is defined pictorially in Fig.~\ref{fig:ssptham} (bottom), and $\widetilde{B}_c=X_c U_c$ with
\begin{equation} \label{eq:uc}
U_c=\prod_{f\in c}U_f 
\end{equation}
where $f\in c$ runs over the six faces of $c$, $U_f$ is a product of four $CZ$ operators in a diamond on face $f$, see Fig.~\ref{fig:ssptham}.

$H_{SSPT}$ has a unique ground state which we denote by $|SSPT\rangle$. We can get some intuition for this ground state using the viewpoint of decorated domain walls (DDW) \cite{Chen2014}, as shown in Fig.~\ref{fig:clustersoup}. Let $\mathcal{C}\subset C$ be a set of body center qubits, and define the state $|\mathcal{C}\rangle$ such that each qubit in $\mathcal{C}$ is in the $|1\rangle$ state, while the rest are in $|0\rangle$. Noting that $CCZ|i\rangle |j\rangle |k\rangle= |i\rangle \otimes CZ^i|j\rangle |k\rangle$, we can write the ground state in the following way,
\begin{equation} \label{eq:ddw}
|SSPT\rangle=\sum_{\mathcal{C}\subset C} |\mathcal{C}\rangle\otimes |\mathcal{G}_\mathcal{C}\rangle
\end{equation}
where the sum is over all subsets $\mathcal{C}$ of $C$ and we have defined the state $|\mathcal{G}_\mathcal{C}\rangle$ on the edge qubits as,
\begin{equation}
|\mathcal{G}_\mathcal{C}\rangle=\prod_{c\in\mathcal{C}} U_c \, |+\rangle^{\otimes |E|} \ .
\end{equation}
Since the $CZ$ gates cancel out between neighboring cubes in $\mathcal{C}$, $|\mathcal{G}_\mathcal{C}\rangle$ describes a state of 2D cluster states on the domain walls of $\mathcal{C}$, and $|+\rangle$ states away from them. The 2D cluster state is the paradigmatic example of a state with SSPT order under line-like symmetries \cite{Raussendorf2019,You2018}. Therefore, $|SSPT\rangle$ can be described as an equal weight superposition over all body qubit configurations in which the domain walls of the body qubits are decorated with 2D SSPT states. 

Let us turn to the symmetries of $|SSPT\rangle$. First, by nature of the DDW structure of $|SSPT\rangle$, we have a global $\mathbb{Z}_2$ symmetry acting on the body qubits $X_C=\prod_{c\in C} X_c$, which follows simply from the fact that the domain walls are invariant under flipping all body spins. The edge qubits have planar subsystem symmetries. For any plane moving parallel to one of the coordinate planes of the cubic lattice, we define the subset $\mathcal{P}\subset E$ as the set of edges that are intersected by this plane. We remark that these planes come in two distinct types, determined by whether they are parallel or perpendicular to the edges they intersect, and we refer to the two types as lattice-planes and dual-planes, respectively. We then define the subsystem symmetry for each plane as $X_\mathcal{P}=\prod_{e\in\mathcal{P}} X_e$. The fact that $X_\mathcal{P}$ is a symmetry of $|SSPT\rangle$ for all planes $\mathcal{P}$ can be seen by first noticing that $[U_c,X_\mathcal{P}]=0$ for all $c\in C$ and all $\mathcal{P}$. Then it follows that $X_\mathcal{P}$ is a symmetry of $|\mathcal{G}_\mathcal{C}\rangle$ for all $\mathcal{C}\subset C$, and therefore is also a symmetry of $|SSPT\rangle$ thanks to Eq.~(\ref{eq:ddw}). 

There is a more insightful way to understand the presence of the planar symmetries. Observe that the intersection of a plane $\mathcal{P}$ with a domain wall configuration forms closed 1D loops, as pictured in Fig.~\ref{fig:leadb}. Since $X_\mathcal{P}$ acts trivially away from the domain walls, we can restrict the action of $X_\mathcal{P}$  onto these closed loops. This is a symmetry, since the 2D cluster states living on the domain walls have line-like subsystem symmetries. Therefore, the planar symmetries of the 3D SSPT follow from the line-like symmetries of the 2D SSPT states which we use to decorate domain walls \footnote{We note that this picture breaks down when $\mathcal{P}$ is tangent to the domain walls, in which case $X_\mathcal{P}$  rather acts as a patch of global symmetry, which turns out to still be a symmetry}. 

We can see explicitly that $|SSPT\rangle$ is in a trivial phase if only the global symmetry or only the subsystem symmetries are enforced by constructing disentangling circuits which respect one of the symmetries. Namely, if we group all of the $CCZ$ gates from $U_{CCZ}$ that live in a given cube, we obtain a local unitary that respects all of the subsystem symmetries. Applying this unitary to all cubes is therefore a subsystem symmetry-respecting circuit that disentangles $|SSPT\rangle$. A disentangling circuit which respects the global symmetries can be obtained by similarly grouping $CCZ$s into octahedrons, one for each face of the lattice. Importantly, neither disentangling circuit respects both types of symmetry, and we argue in the next section that such a symmetric disentangling circuit does not exist by virtue of the non-trivial SSPT order.

\subsection{Boundary of the SSPT}

\begin{figure}
\centering
\includegraphics[width=\linewidth]{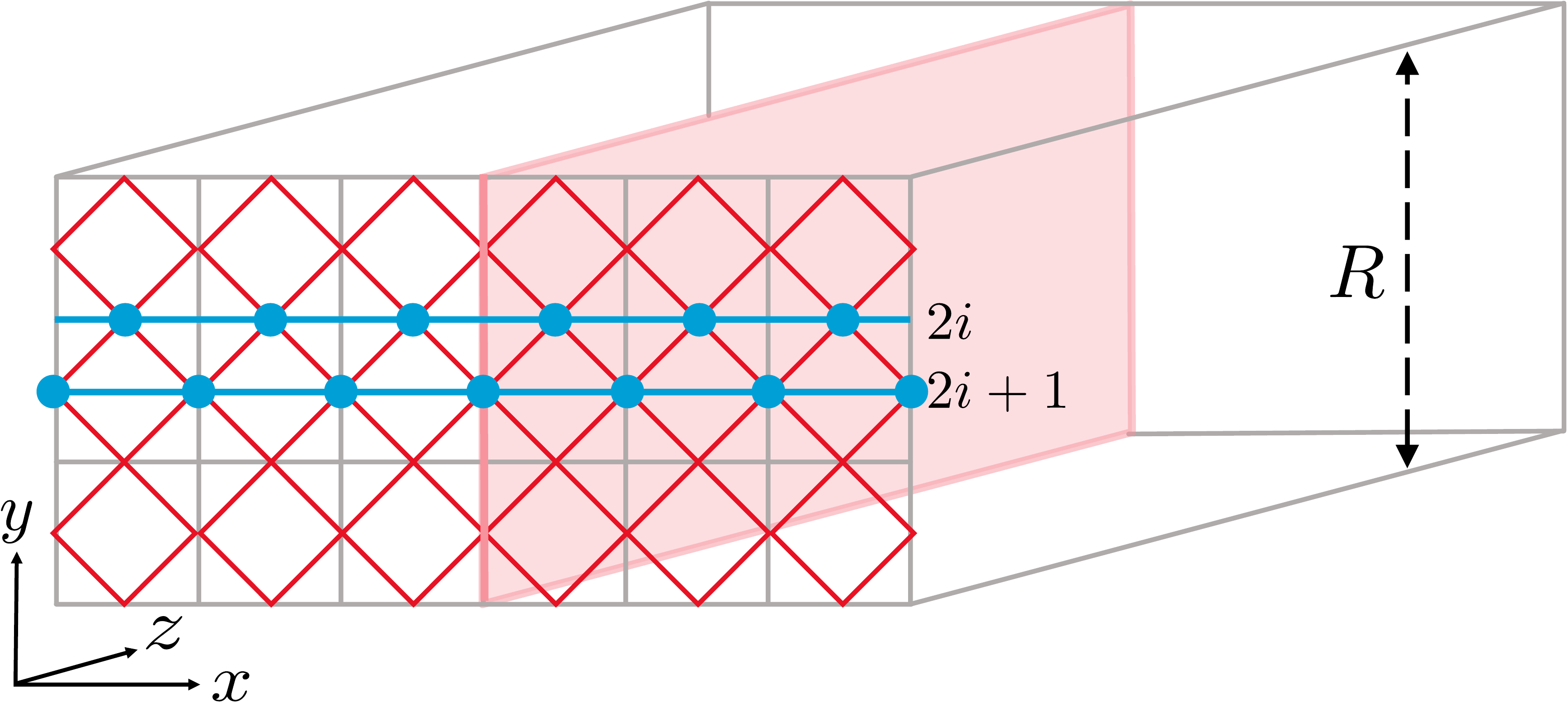}
\caption{The 3D cylinder with periodic boundary conditions in the $x$ and $y$ directions, and open boundaries in the $z$ direction. The effective action of the bulk symmetries on the boundary is shown. The red lines denote CZ gates coming from the global symmetry action, while blue lines indicate the action of $xz$ planar symmetries, with blue dots denoting $X$ operators. The plane in the $yz$ direction represents a domain wall that is inserted to create a global symmetry flux through the cylinder.}
\label{fig:boundary}
\end{figure}

We now demonstrate the non-trivial nature of our SSPT model by analyzing its boundary. Let us consider the geometry pictured in Fig.~\ref{fig:boundary}, where the boundary is a 2D square lattice with periodic boundary conditions and qubits on the edges. In the presence of this boundary, the whole 3D state may no longer be symmetric under the global or subsystem symmetries. In particular, it may be necessary to dress symmetry operators with additional action on the boundary qubits in order to leave the system invariant. For the planar symmetries, this turns out to be unnecessary, and the action of the planar symmetries on the boundary, for planes perpendicular to the boundary, is simply a line of $X$'s. The global symmetry, on the other hand, must be decorated with additional $CZ$'s acting between every nearest neighboring pair of edges on the boundary. The action of the symmetries on the boundary is summarized in Fig.~\ref{fig:boundary}.

We see from the above analysis that neighbouring plane symmetries commute on the boundary. This is different from the 2D SSPT order of the 2D cluster state, where neighbouring line symmetries commute in the bulk but anti-commute on the boundary. However, we will see that, in the presence of a global symmetry flux threading the cylinder, which is introduced by adding the domain wall pictured in Fig.~\ref{fig:boundary}, neighbouring plane symmetries will anti-commute. This highlights the importance of both global and subsystem symmetries in our SSPT model. 

To formalize this, we can use the language of group cohomology which classifies SPT order. Imagine  ``compactifying'' our 3D system into a quasi-2D system~\cite{Dua2019}. This is achieved simply by fixing the length $R$ in the $y$ direction, as indicated in Fig.~\ref{fig:boundary}. We then get a different quasi-2D system for each compactification radius $R$. The subsystem symmetries perpendicular to the compactification direction, \textit{i.e.} $xz$ planes, become standard global symmetries of the compactified system. There are $2R$ of these symmetries, labelled by indicies for $i=1,\dots,2R$ where even (odd) $i$ corresponds to lattice-plane (dual-plane) symmetries, as pictured in Fig.~\ref{fig:boundary}. Including the global symmetry as well, we can consider the symmetry group,
\begin{align}
G_R&=\mathbb{Z}_2\times\mathbb{Z}_2^{2R} \nonumber \\
&=\left\{\left(g,\vec{s}\right)|\vec{s}=\{s_1,\dots,s_{2R}\}\, ;\, g,s_i=0,1\right\},
\end{align}
where the generators $g$ and $s_{i}$ correspond to the global and subsystem symmetries, respectively.

For each $R$, we can study the 2D SPT order of the compactified system under the symmetry group $G_R$. This order is determined by a 3-cocycle $\omega:G_R\times G_R\times G_R\rightarrow U(1)$ corresponding to a cohomology class $[\omega]\in H^3(G,U(1))$ which characterizes the action of $G_R$ on the boundary \cite{Chen2011,Zalatel2014}. We can straightforwardly calculate this cocycle using the procedure introduced in Ref.~\cite{Else2014}. The result is,
\begin{equation}
\omega\left((g,\vec{s}),(g',{\vec{s}\,}'),(g'',{\vec{s}\,}'')\right)=(-1)^{g''\sum_{i=1}^R s'_{2i}(s_{2i+1}+s_{2i-1})} .
\end{equation}
We can determine the effect of inserting a global symmetry flux using the slant product, as described in Refs.~\cite{Zalatel2014,williamson2014matrix}. The slant product corresponding to a group element $a$ is a function $\chi_a:G_R\times G_R\rightarrow U(1)$ defined as,
\begin{equation} 
\chi_a(b,c)=\frac{\omega(a,b,c)\omega(b,c,a)}{\omega(b,a,c)}  .
\end{equation}
When $\omega$ is a 3-cocycle, $\chi_a$ will be a 2-cocycle, corresponding to a cohomology class $[\chi_a]\in H^2(G,U(1))$. The physical meaning of the slant product is the following. For any $a\in G$, let $V_a(b)$ be the action of $b\in G$ on the boundary in the presence of an $a$-flux. This action may be projective, \textit{i.e.} $V_a(b)V_a(c)$ may equal $V_a(bc)$ only up to a phase. This phase is precisely given by the slant product, $V_a(b)V_a(c)=\chi_a(b,c) V_a(bc)$. In the case that $\chi_a$ belongs to a non-trivial cohomology class in $H^2(G,U(1))$, the $a$-flux carries a non-trivial 1D SPT order. We remark that, for finite abelian $G$, $\chi_a$ is in a trivial cohomology class if and only if it is symmetric, \textit{i.e.} $\chi_a(b,c)=\chi_a(c,b)$ \cite{Kleppner1965}. Since $V_a(b)V_a(c)=\frac{\chi_a(b,c)}{\chi_a(c,b)}V_a(c)V_a(b)$, the cohomology of $\chi_a$ is trivial if and only if the action of the symmetry on the boundary commutes. Furthermore, a 3-cocycle $\omega$ whose slant product (for some $a$) belongs to a non-trivial class in $H^2(G,U(1))$ must itself belong to a non-trivial class of $H^3(G,U(1))$.

If we compute the slant product with $a=(1,\vec{0})$ corresponding to inserting a global symmetry flux via the domain wall added in Fig.~\ref{fig:boundary}, we find,
\begin{equation} \label{eq:slant}
\chi_{(1,\vec{0})}\left( (g,\vec{s}),(g',{\vec{s}\,}')\right)=(-1)^{\sum_{i=1}^R s'_{2i}(s_{2i+1}+s_{2i-1})}
\end{equation}
In particular, the commutation relation reads,
\begin{equation} \label{eq:slantcomm}
\frac{\chi_{(1,\vec{0})}\left( (g,\vec{s}),(g',{\vec{s}\,}')\right)}{\chi_{(1,\vec{0})}\left( (g',{\vec{s}\,}'),(g,\vec{s})\right)}=(-1)^{\sum_{i=1}^{2R} \left(s_is'_{i+1}+s_i' s_{i+1}\right)}
\end{equation}
This says precisely that neighbouring plane symmetries anti-commute on the boundary in the presence of a global symmetry flux. We can also compute the slant product for a subsystem symmetry flux, corresponding to $a=(0,\hat{i})$, where the vector $\hat{i}$ contains 1 at position $i$, and 0 elsewhere. We find,
\begin{equation} \label{eq:subslant}
\chi_{(0,\hat{i})}\left( (g,\vec{s}),(g',{\vec{s}\,}')\right)=(-1)^{g'(s_{i-1}+s_{i+1})},
\end{equation}
which tells us that, in the presence of a lattice-plane (dual-plane) subsystem symmetry flux, the global symmetry anti-commutes on the boundary with the two dual-plane (lattice-plane) subsystem symmetries neighbouring the flux.

One might wonder if we are missing out on any important information by only considering subsystem symmetries in one direction (the $xz$ planar symmetries). Indeed, it could be the case that, \textit{e.g.} perpendicular symmetry planes anti-commute in the presence of a certain symmetry flux. However, we have checked that this is not the case: the cases considered above capture all of the non-trivial fractionalization that occurs in our model. 

From the non-trivial slant products computed above, we see that the 3-cocycle $\omega$ belongs to a non-trivial cohomology class for all compactification radii $R$. The general arguments of Ref.~\cite{Chen2011} then show that the boundary, when considered as a quasi-1D system, cannot be gapped and symmetric; either the symmetry will be spontaneously broken, resulting in a boundary degeneracy, or the boundary will be gapless. 
In addition to this, when we enforce a true 2D notion of locality, the boundary could potentially gain even more non-trivial features that we miss by employing compactification. In Appendix \ref{app:boundary}, we examine some possible Hamiltonians which respect the boundary symmetries. We find that, unlike in conventional SPT phases, neither of the simplest choices of boundary Hamiltonian lead to a gapless boundary, suggesting that it may be necessary to look beyond this compactified description to fully understand this boundary. We leave such an analysis to future work.

\section{Gauging the global symmetry: Subsystem symmetry enriched topological order} \label{sec:sset}

In this section, we gauge the global $\mathbb{Z}_2$ symmetry of $|SSPT\rangle$. The resulting model is a $\mathbb{Z}_2$ gauge theory in which the loop-like topological excitations fractionalize under the subsystem symmetries. We therefore call this model an example of SSET order. We show that this implies an extensive degeneracy of the loop excitations which is protected by the symmetry. We also show that the model has an enlarged value of the topological entanglement entropy, as compared to the underlying topological order, due to the subsystem symmetry enrichment. 

Let us briefly describe the gauging procedure we which we employ (more details may be found in Refs.~\cite{Vijay2016,Williamson2016,Shirley2019}). The gauging procedure maps body qubits to qubits on the faces ($F$) of the lattice, in such a way that the face qubits $f\in F$ take the state $|1\rangle$ on the domain walls of the body qubits, and $|0\rangle$ elsewhere. That is, given a state $|\mathcal{C}\rangle$ on the body qubits, the effect of the gauging map $\Gamma$ can be written as $\Gamma|\mathcal{C}\rangle=|\partial \mathcal{C}\rangle$, where $\partial \mathcal{C}\subset F$ is the set of faces on the boundary of $\mathcal{C}$, and $|\partial \mathcal{C}\rangle$ describes a state on the face qubits where all qubits in $\partial \mathcal{C}$ are in the state $|1\rangle$, and the rest are in $|0\rangle$. We note that the edge qubits are unaffected under the action of this map. We can then extend the map $\Gamma$ to arbitrary states on the body qubits by linearity. Applying this procedure to $|SSPT\rangle$, we get,
\begin{align} \label{eq:sset}
|SSET\rangle&=\Gamma|SSPT\rangle \nonumber \\
&=\sum_{\mathcal{C}\subset C} \Gamma|\mathcal{C}\rangle\otimes |\mathcal{G}_\mathcal{C}\rangle \nonumber \\
&=\sum_{\mathcal{C}\subset C} |\partial \mathcal{C}\rangle \otimes |\mathcal{G}_\mathcal{C}\rangle
\end{align}
This state can again be visualized using Fig.~\ref{fig:clustersoup}, where now the colored faces indicate elements of $\partial\mathcal{C}$, which can be viewed as a configuration of closed membranes on the faces of the lattice. Therefore, $|SSET\rangle$ is a superposition over all closed membranes configurations on the face qubits, where the membranes are decorated with 2D cluster states on the edge qubits. It is clear that $|SSET\rangle$ has the same subsystem symmetries on the edge qubits as $|SSPT\rangle$.

If we were to remove $|\mathcal{G}_\mathcal{C}\rangle$ from the above equation, the state described would be exactly the 3D toric code, $|TC\rangle=\sum_{\mathcal{C}\subset C} |\partial\mathcal{C} \rangle$, which has topological order. In fact, we can disentangle the edge qubits from the face qubits using a unitary circuit $\widetilde{U}_{CCZ}$ which places four $CCZ$'s on each face as in Fig.~\ref{fig:ssetham}.
Applying this circuit to $|SSET\rangle$, we get,
\begin{align}
\widetilde{U}_{CCZ}|SSET\rangle &=\sum_{\mathcal{C}\subset C} \tilde{U}_{CCZ}(|\partial\mathcal{C}\rangle)\otimes |\mathcal{G}_\mathcal{C}\rangle) \\
&= \sum_{\mathcal{C}\subset C} |\partial\mathcal{C}\rangle\otimes \prod_{c\in\mathcal{C}} U_c\, |\mathcal{G}_\mathcal{C}\rangle \\
&=|TC\rangle\otimes |+\rangle^{\otimes |E|}
\end{align}
Therefore, $|SSET\rangle$ is related to $|TC\rangle$ by a unitary circuit, up to trivial degrees of freedom, so the two states have the same topological order. However, this circuit does not respect the subsystem symmetries. In fact, we will show that, when these symmetries are enforced, $|SSET\rangle$ is in a distinct phase from $|TC\rangle$, as indicated by symmetry enrichment of the topological excitations. Therefore, we can say that $|SSET\rangle$ has subsystem symmetry enriched topological order.

\subsection{Excitations and symmetry enrichment}

\begin{figure}[t]
\centering
\subfigure[]{\label{fig:ssetham}\includegraphics[scale=0.08]{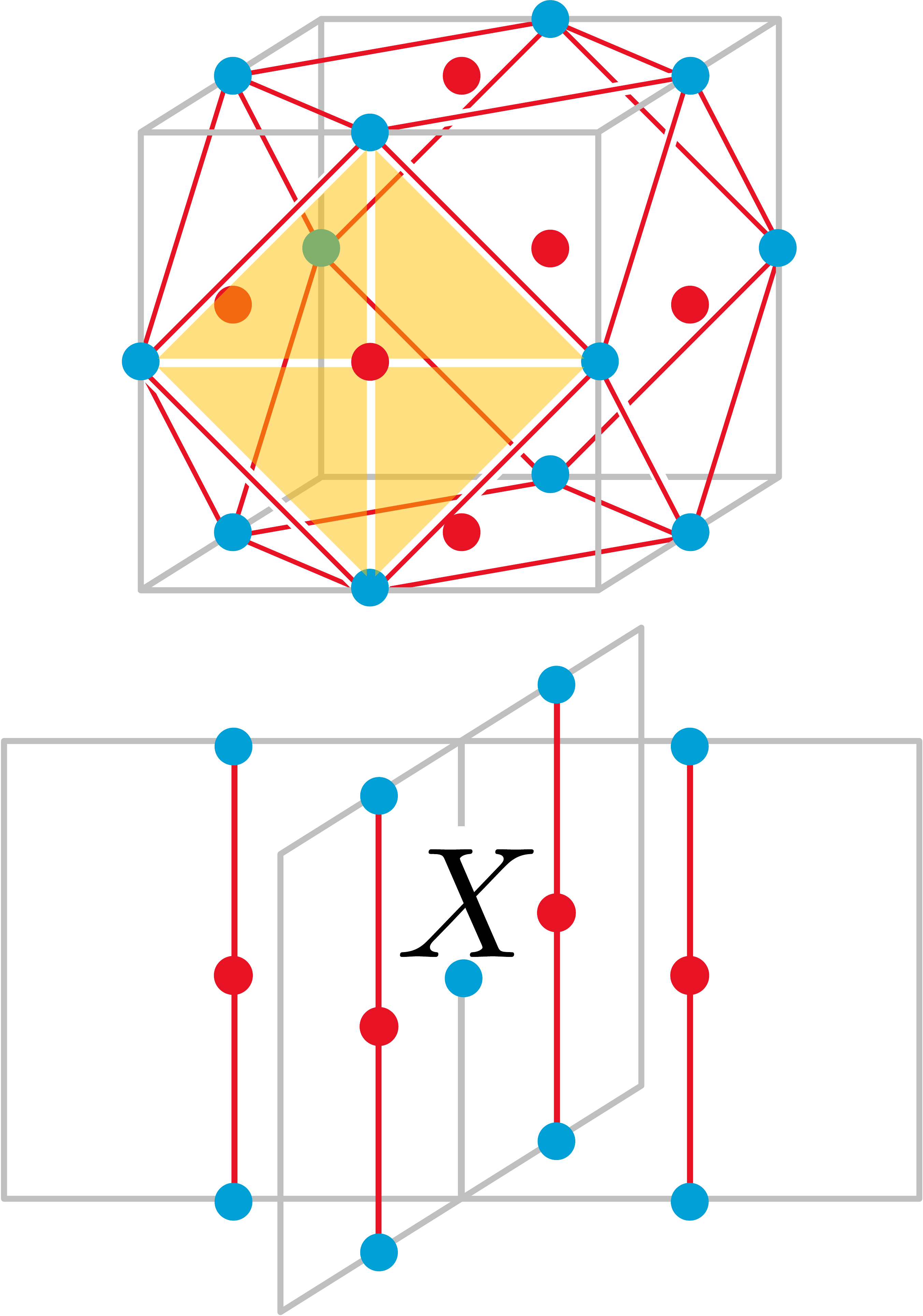}} \hfill
\subfigure[]{\label{fig:3dexcitations}\includegraphics[scale=0.08]{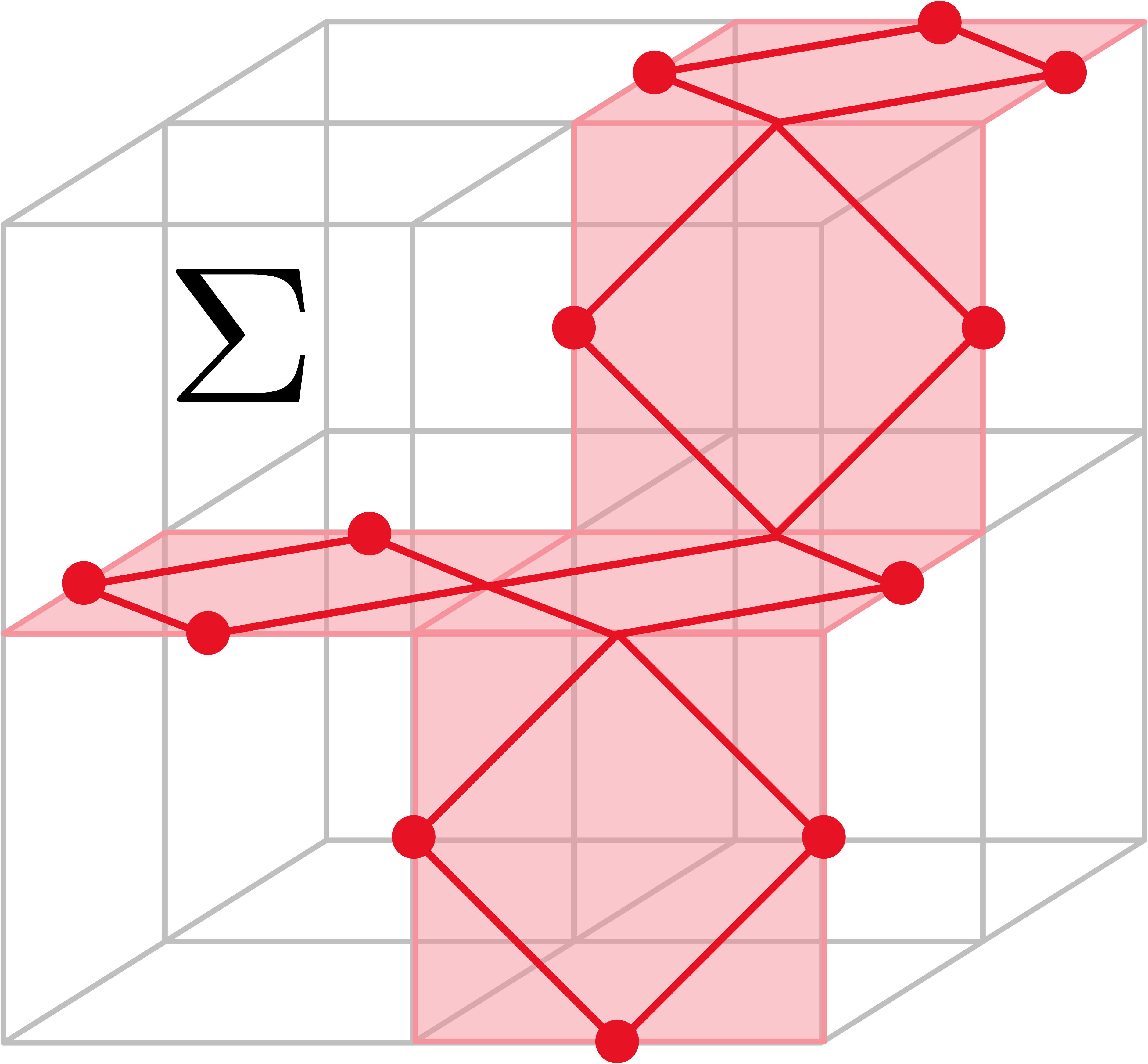}} \hfill
\subfigure[]{\label{fig:2dcut}\includegraphics[scale=0.13]{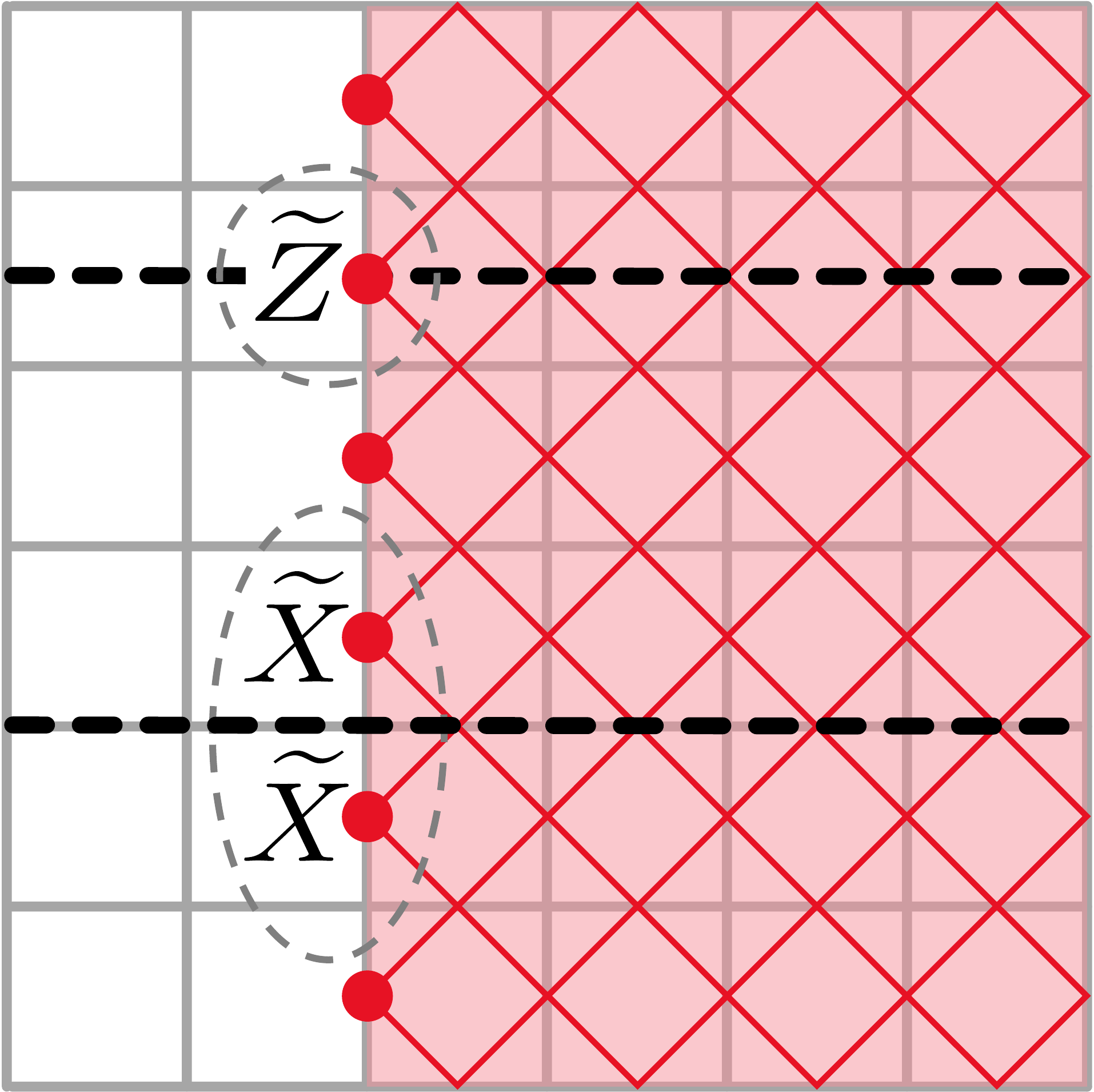}}
\caption{(a) Top: a unit cell of the 3D SSET state. Four of the triangles $\triangle$ appearing in $\widetilde{U}_{CCZ}$ are shown; there are four such triangles on each face of the lattice. Bottom: The Hamiltonian term $C_e$, where $e$ is the central vertical edge. (b) A loop-like excitation appearing on the boundary of the surface $\Sigma$, indicated by shaded faces. Excitations appear on edges marked by points. (c) A 2D cut of our 3D lattice with a loop excitation corresponding to the half-infinite membrane $\Sigma^\infty$, as viewed from above. 
There is an effective qubit degree of freedom for each edge on the boundary of $\Sigma^\infty$, indicated by dots. The horizontal dashed lines represent the intersection of two symmetry planes $\mathcal{P}$ with this 2D slice, and the action of $X_\mathcal{P}$ on the edge degrees of freedom, according to Eq.~\ref{eq:loopsymmaction}, is shown beside each line.}
\end{figure}

The Hamiltonian for which $|SSET\rangle$ is a ground state can be obtained by gauging $H_{SSPT}$, and has the following form,
\begin{equation}
H_{SSET}=-\sum_{e\in E} A_e -\sum_{c\in C} B_c-\sum_{e\in E} C_e\frac{1+A_e}{2}
\end{equation}
where,
\begin{align}
&A_e=\prod_{f\ni e} Z_f \\
&B_c= U_c \prod_{f\in c} X_f
\end{align}
where $f\ni e$ denotes all faces incident on edge $e$. $C_e$ is defined pictorially in Fig.~\ref{fig:ssetham}. The new gauge term $A_e$ enforces that the faces qubits form closed membranes in the ground state. The modified terms $B_c$ and $C_e$ are simply the original terms $\widetilde{B}_c$ and $\widetilde{C}_e$ rewritten as functions of the new face qubits, using the rules
$X_c\mapsto \prod_{f\in c} X_f$ and $\prod_{c\ni f} Z_c\mapsto Z_f$
that follow from the gauging procedure. In addition, we project the $C_e$ term onto the closed membrane subspace ($A_e=1$ $\forall e\in E$) in order to ensure that the Hamiltonian respects the subsystem symmetries. We note that all terms in $H_{SSET}$ commute. On a topologically non-trivial manifold like a three-dimensional torus, the ground space of $H_{SSET}$ is degenerate. $|SSET\rangle$ is one of the degenerate ground states, and the others can be obtained by adding non-contractible membranes into the superposition in Eq.~(\ref{eq:sset}).

Let us now construct the excitations of $H_{SSET}$, which come in three types. Violations of $C_e$ are topologically trivial particles that can be created locally by acting with $Z_e$, so we do not discuss them further. The other two types of excitation are topologically non-trivial, and must be created by extended non-local operators. Violations of $B_c$, called electric excitations, are point-like, and are created in pairs at the end points of string operators,
\begin{equation}
S^\mathrm{e}_\Lambda=\prod_{f\in\Lambda} Z_f
\end{equation}
where $\Lambda\subset F$ is a curve on the dual lattice that penetrates faces. Violations of $A_e$, called magnetic excitations, are loop-like. They appear along the boundary of open membrane operators,
\begin{equation}
S^\mathrm{m}_\Sigma = \prod_{f\in \Sigma} X_f U_f,
\end{equation}
where $\Sigma\subset F$ is an open membrane of faces, as pictured in Fig.~\ref{fig:3dexcitations}. If we were to braid an electric excitation through the loop of a magnetic excitation, we would pick up a minus sign due to the anti-commutation of $X$ and $Z$. This is the same as in the toric code. 

The difference from the toric code comes from the degeneracy of the excitations. In particular, since $A_e = -1$ when acting on the boundary of $\Sigma$, the projection $(1+A_e)/2$ removes the corresponding terms $C_e$, so we can decorate the boundary of the membrane operator $S^\mathrm{m}_\Sigma$ with $Z$ operators without changing the energy of the resulting excitation:
\begin{equation}
S^\mathrm{m}_{\Sigma}(\{a_e\}_{e\in\partial\Sigma})= S^\mathrm{m}_{\Sigma} \prod_{e\in\partial\Sigma} Z_e^{a_e},\quad a_e=0,1, 
\end{equation}
where $\partial\Sigma$ denotes all edges on the boundary of the membrane $\Sigma$. Thus the loop-like excitation is two-fold degenerate per unit length. This degeneracy is protected by the subsystem symmetries. Intuitively, this is because the loop-like excitation coincides with the boundary of a 2D cluster state which has an exponential boundary degeneracy protected by the subsystem symmetries.

More precisely, consider a membrane $\Sigma^\infty$ which is a half-infinite plane, as shown in Fig.~\ref{fig:2dcut}. This creates an excitation lying along $\partial\Sigma^\infty$ which is a line of $N$ edges that we denote by $e_i$ for $i=1,\dots N$. The degenerate subspace associated to this excitation is spanned by the states $|a_1,\dots a_n\rangle$ defined as,
\begin{equation} \label{eq:loopedgebasis}
|a_1,\dots a_n\rangle:=S^\mathrm{m}_{\Sigma^\infty} \prod_{i=1}^N Z_{e_i}^{a_i}\, |SSET\rangle\ , \quad a_i=0,1\ .
\end{equation}
Let us now determine the action of the subsystem symmetries in this $N$-qubit space. Consider those symmetry planes that are perpendicular to $\Sigma^\infty$ and also cross $\partial\Sigma^\infty$, as pictured in Fig.~\ref{fig:2dcut}. We can index these planes by $\mathcal{P}_i$ and $\mathcal{P}_{i+1/2}$, corresponding to dual-planes intersecting edge $e_i$, or lattice-planes intersecting between edges $e_i$ and $e_{i+1}$, respectively. Then we can calculate,
\begin{align} \label{eq:loopsymmaction}
&X_{\mathcal{P}_i}|a_1,\dots a_n\rangle=(-1)^{a_i} |a_1,\dots a_n\rangle \\
&X_{\mathcal{P}_{i+1/2}}|a_1,\dots a_n\rangle = |a_1,\dots, a_i\oplus 1,a_{i+1}\oplus 1,\dots, a_n\rangle. \nonumber
\end{align}
Therefore, if we let $\widetilde{X}_i,\widetilde{Z}_i$ denote the logical Pauli operators in the degenerate edge subspace, we have $X_{\mathcal{P}_i}\cong \widetilde{Z}_i$ and $X_{\mathcal{P}_{i+1/2}}\cong \widetilde{X}_i\widetilde{X}_{i+1}$. We see that neighboring plane symmetries anticommute on the edge of the excitation. This is the same pattern of symmetry fractionalization found on the boundary of the 2D cluster state \cite{You2018}. In the case the case of the 2D cluster state, the line-like subsystem symmetry protects the exponential edge degeneracy of the 2D cluster state. In analogy, the planar subsystem symmetry here protects the exponential degeneracy of the line-like excitation. A similar discussion holds for an arbitrary membrane $\Sigma$, although there can be some finite size effects due to corners, as discussed in Ref.~\cite{You2018}. Therefore, the loop-like magnetic excitations of $|SSET\rangle$ carry a two-fold degeneracy per unit length that is protected by the planar subsystem symmetries. 

\subsection{Topological entanglement entropy}

In this section, we show that the topological entanglement entropy (TEE) of our 3D SSET state is larger than that of the 3D toric code when the boundaries of bipartitions are aligned with the symmetry planes. This is in analogy to the fact that states with 2D SSPT order, such as the cluster state, have a non-zero TEE when biparitions are aligned with the line-like symmetries, despite the absence of topological order \cite{Zou2016,Williamson2019,Stephen2019,Devakul2018}. To emphasize that this value comes from the SSPT order, rather than topological order, we call it the symmetry-protected entanglement entropy (SPEE) \cite{Stephen2019}. In this case, we say that the 3D SSET has a non-zero SPEE in addition to the TEE inherited from the topological order of the 3D toric code.

The origin of the SPEE can be imagined as follows. Consider one closed membrane in the membrane soup defining $|SSET\rangle$. If we bisect the membrane with a plane aligned with the subsystem symmetries, as in Fig.~\ref{fig:lead}, we can calculate the entanglement between the two halves of the membrane. Since the membranes are decorated by 2D cluster states, and the cut is aligned with the line-like symmetries of the cluster state, we find a non-zero SPEE for this membrane. Since $|SSET\rangle$ is a fluctuating soup of such membranes, one can imagine that it also exhibits a non-zero SPEE. In the rest of this section, we show that this is indeed the case, with details of the calculation presented in Appendix~\ref{app:tee}.

\begin{figure}[t]
\centering
\subfigure[]{\label{fig:3torus}\includegraphics[scale=0.125]{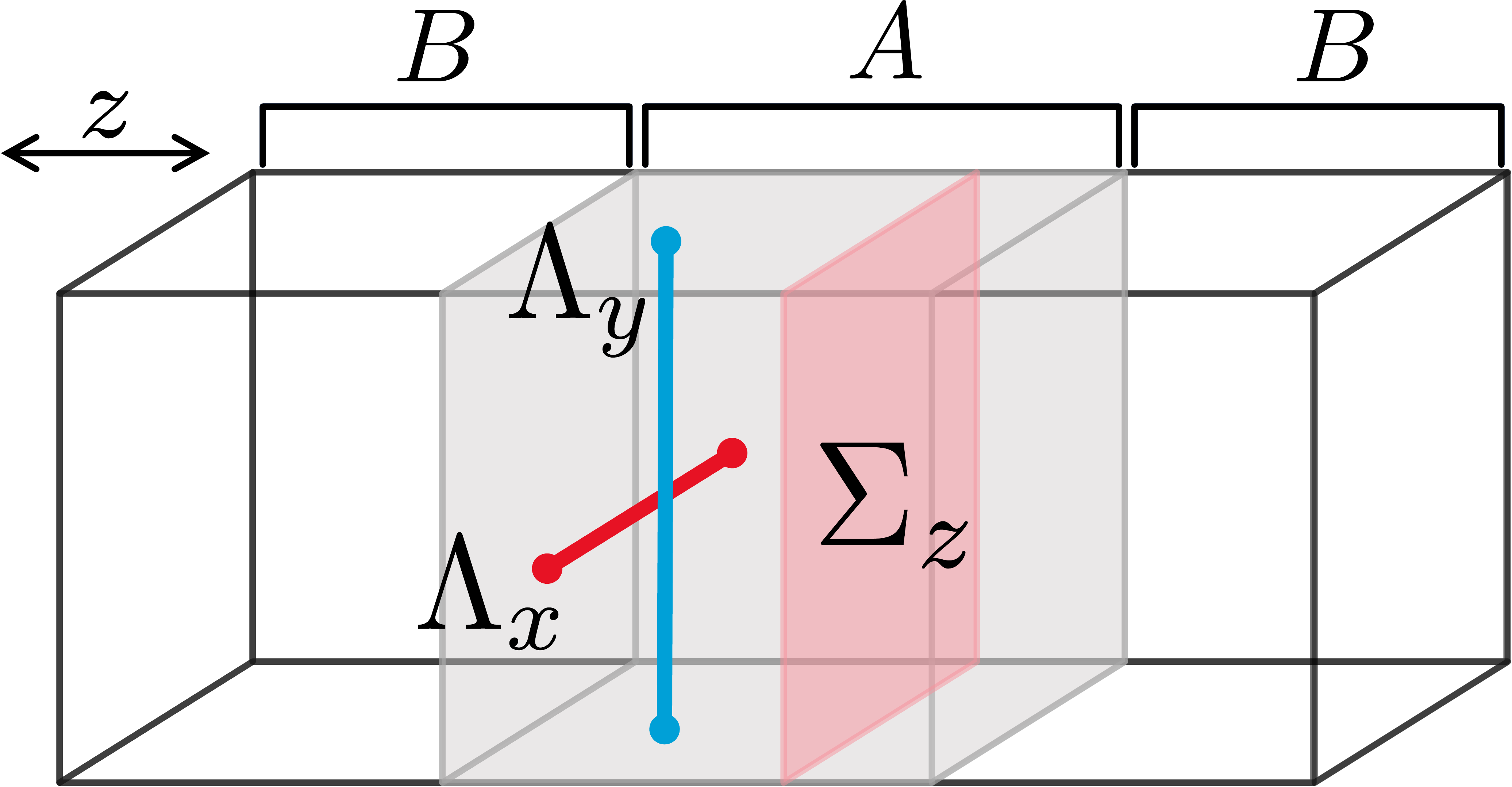}} \hfill
\addtocounter{subfigure}{1}
\subfigure[]{\label{fig:frame}\includegraphics[scale=0.115]{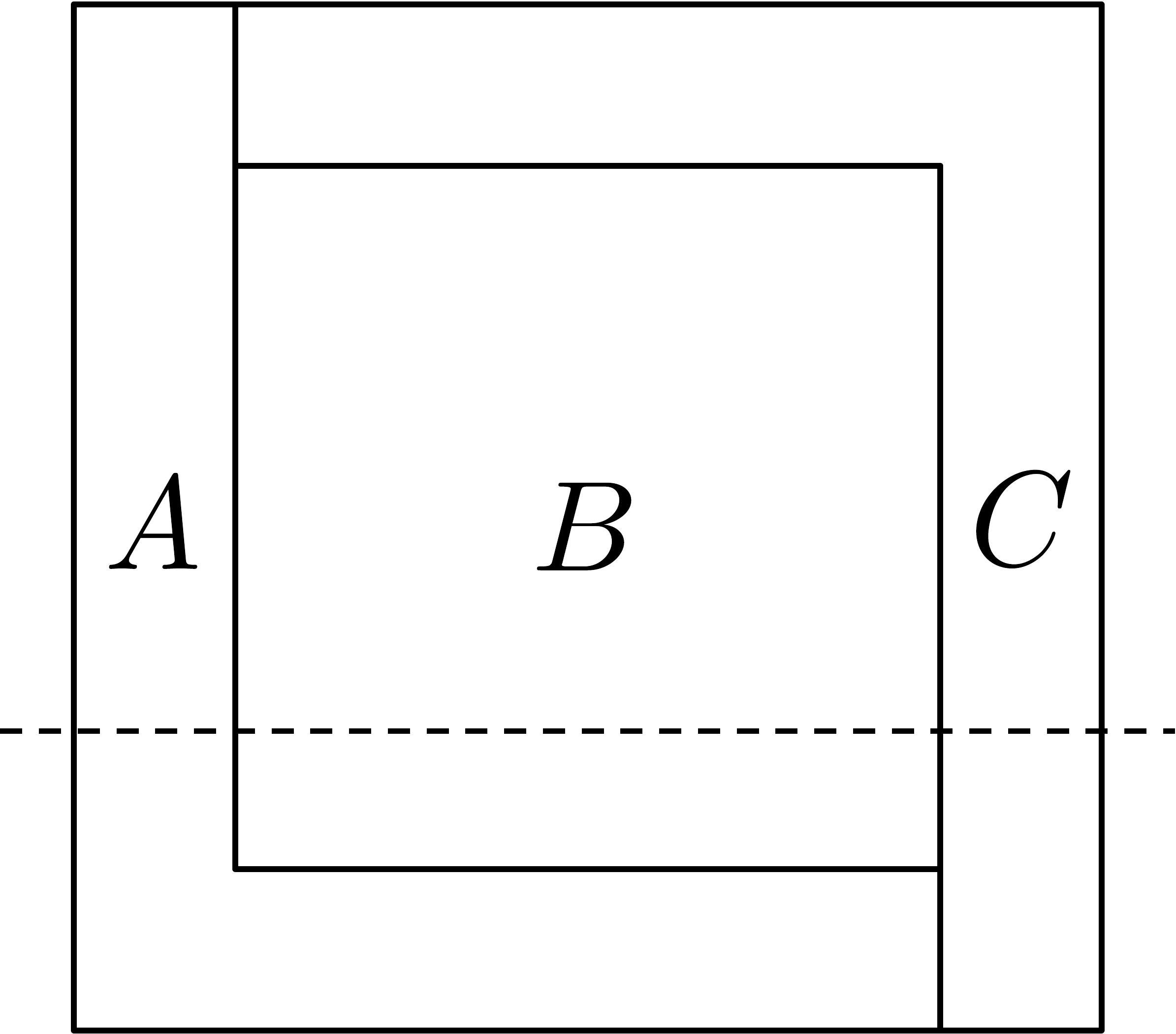}}  \\
\hspace{3.5mm}
\addtocounter{subfigure}{-2}
\subfigure[]{\raisebox{6mm}{\label{fig:dumb}\includegraphics[scale=0.115]{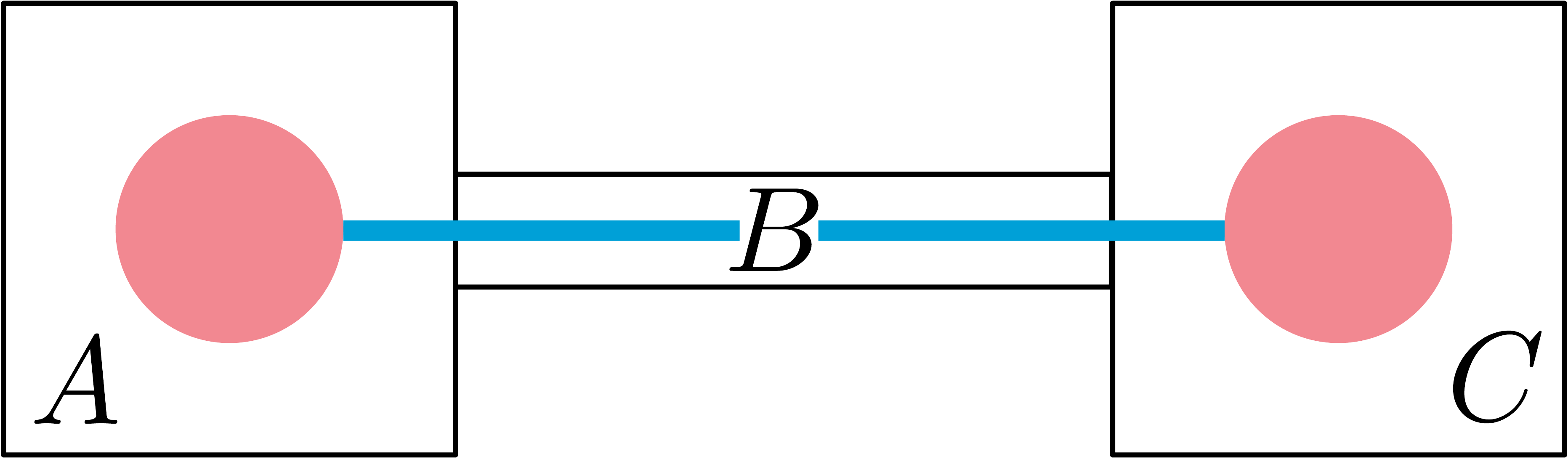}}}\hfill
\addtocounter{subfigure}{1}
\subfigure[]{\label{fig:frame2}\includegraphics[scale=0.115]{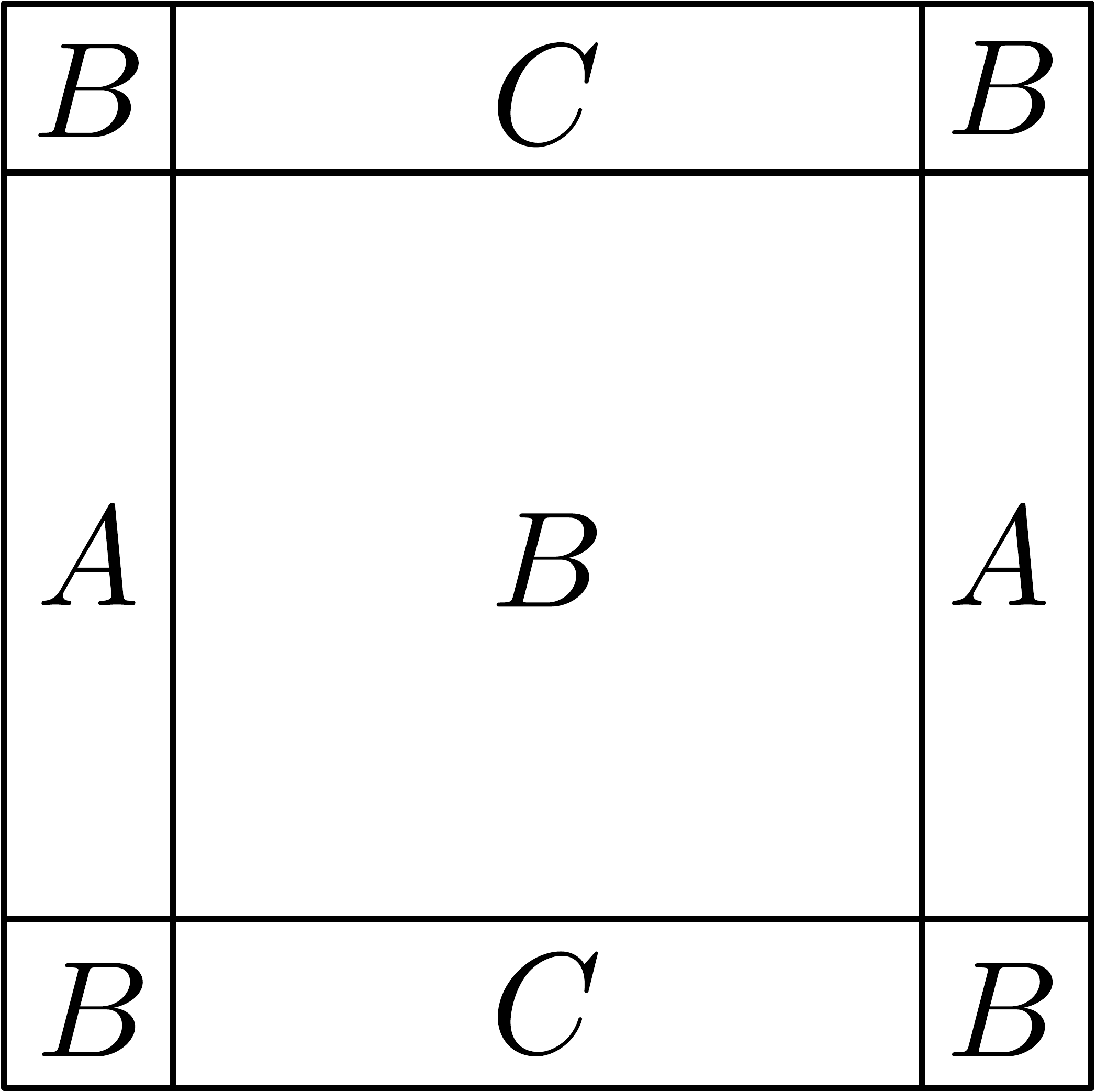}} 
\hspace{0.9 mm}
\caption{(a) The geometry considered throughout this section. Opposite sides are identified, resulting in a 3D torus. The $A$ and $B$ subsystems are shown, as well as the non-contractible loops $\Lambda_{x,y}$ and membrane $\Sigma_{z}$.  (b) The dumbbell configuration for detecting the SPEE in 2D systems from Ref.~\cite{Williamson2019}. (c) The picture frame configuration proposed to detect the SPEE that arises due to planar subsystem symmetries in 3D systems. Here we see a view from the top. A cross section along the dotted line reveals a shape identical to that of (b). (d) The alternate picture frame configuration that does not detect SPEE due to line-like symmetries.}
\end{figure}

For simplicity we calculate the entropy of a finite section of a 3D torus, such that the boundary between subsystems $A$ and $B$ is two disconnected 2D tori. However, we expect the same results to hold for any geometry that is appropriately aligned with the subsystem symmetries, as in Ref.~\cite{Williamson2019}. We will aim to determine the 2-R\'enyi entropy $S^{(2)}_A=-\ln \mathrm{Tr}(\rho_A^2)$ where $\rho_A$ is the reduced state of subsystem $A$ \footnote{Ideally, we would calculate the Von-Neumann entropy, but this is more difficult since the entanglement spectrum will turn out to be not flat}. On a 3D torus, $H_{SSET}$ has eight degenerate ground states. We choose $|SSET\rangle$ to be one of the minimal entropy states which have the largest TEE~\cite{Zhang2012}. This is done by picking the ground state that is +1 eigenstate of the membrane operators $S^\mathrm{m}_{\Sigma_{z}}$ where  $\Sigma_{z}$ is the non-contractible membrane , as well as the loop operators $S^{\mathrm{e}}_{\Lambda_{x/y}}$ where $\Lambda_{x/y}$ are the non-contractible loops, as pictured in Fig.~\ref{fig:3torus}. In this way, the subsystem $A$ has maximum knowledge of the whole state, and hence minimum entropy. We define $G$ to be the abelian group generated by all Hamiltonian terms $A_e$, $B_c$, $C_e$ as well as $S^\mathrm{m}_{\Sigma_{z}}$ and $S^{\mathrm{e}}_{\Lambda_{x/y}}$.

For simplicity, we assume that region $A$ contains $L\times L\times L$ vertices, such that $L^2$ edges are cut on each boundary. In Appendix \ref{app:tee}, we show that the entropy can be expressed in the following way for large $L$,
\begin{equation} \label{eq:entropy}
S^{(2)}_A=(|A|+2L^2) \ln 2-\ln |G_A| - 2\mathcal{F}(\ln\sqrt{2})
\end{equation}
where $|A|=6L^3+3L^2$ is the number of qubits in $A$, $G_A$ is the subgroup of $G$ containing those operators that act non-trivially on $A$ only, and $\mathcal{F}(\beta)$ is the (extensive) free energy of a 2D square lattice Ising model at inverse temperature $\beta$. By counting independent generators in the same way as for the 3D toric code, \cite{Castelnovo2008}, we can compute that $|G_A| = 2^{6L^3-2L^2+4}$. Compared to the toric code case, $G_A$ here contains an additional non-local operator for each boundary, namely the product of $C_e$ for all edges $e$ on the boundary, which is equal to the subsystem symmetry on the corresponding dual plane.

To compute the free energy, we can use Onsager's result \cite{Onsager1944}, which yields $\mathcal{F}(\ln\sqrt{2})=\ln 2(\frac{1}{2}+\ln 2) L^2$ for large $L$ (for more details, see Appendix \ref{app:tee}). Crucially, the temperature $\beta=\ln\sqrt{2}$ lies in the disordered phase of the Ising model, such that the free energy is extensive with no constant term. Putting everything together, the entropy is,
\begin{equation}
S^{(2)}_A=2cL^2-2\gamma_{\mathrm{TEE}}-2\gamma_{\mathrm{SPEE}}+\dots
\end{equation}
where $c=(3-\ln 2)\ln 2$, $\gamma_{\mathrm{TEE}}=\gamma_{\mathrm{SPEE}}=\ln 2$, and the dots represent terms that go to zero as $L$ goes to infinity. We include the factors of 2 to emphasize the fact that $A$ has two disconnected boundaries. $\gamma_{\mathrm{TEE}}$ comes simply from $|G_A|$ in the same way as for the 3D toric code, whereas $\gamma_{\mathrm{SPEE}}$ is due to the subsystem symmetries forming non-local constraints on the boundary of $A$.  

While we have only shown the existence of a non-zero SPEE for this specific bipartition, we expect that it would also be present for other bipartitions whose geometry are aligned with the symmetry planes, as is the case for the SPEE of the 2D cluster state \cite{Williamson2019}. In Ref.~\cite{Williamson2019}, the authors also constructed a method to extract the SPEE due to line-like subsystem symmetries in a 2D system, such that it can be separated from the TEE. They show that the combination of entropies
\begin{equation} \label{eq:sdumb}
S_{\mathrm{SPEE}}= S_B+S_{ABC}-S_{AB}-S_{BC}
\end{equation}
is equal to the SPEE, where subsystems $A,B,C$ form a dumbbell shape as in Fig.~\ref{fig:dumb}. This is due to the fact that the line-like symmetries (blue line in Fig.~\ref{fig:dumb}) can be terminated by applying local operators in the circled red regions, such that we get a constraint that reduces $S_{ABC}$. Crucially, the red regions have a larger radius than the width of $B$, such that the same truncated line cannot fit into $B$, $AB$, or $BC$. On the other hand, the TEE is the same for each of the four terms in Eq.~(\ref{eq:sdumb}), and hence will cancel out, as will all extensive parts of the entropy.

Following the same logic, we propose a set of subsystems $A,B,C$ such that $S_{\mathrm{SPEE}}$ will be equal to the SPEE due to planar symmetries, see Fig.~\ref{fig:frame}. The subsystem $B$ is a thin slab, while $A$ and $C$ are thickened, giving a geometry similar to that of a picture frame. As in the dumbbell, the planar subsystem symmetries can be truncated by applying operators along the boundary. 
Such an operator will fit into region $ABC$ of the frame, but none of the other combinations. Therefore, we conjecture that $S_{\mathrm{SPEE}}$ should be equal to $\gamma_{\mathrm{SPEE}}$ for the picture frame geometry. If there is also a SPEE due to linear subsystem symmetries, then $S_{\mathrm{SPEE}}$ would grow as the perimeter of region $B$. If we take the alternate picture frame geometry in Fig.~\ref{fig:frame2}, we would only detect SPEE due to planar symmetries, since any truncated line operator that contributes to $S_{ABC}$ will also contribute to one of $S_{AB}$ or $S_{BC}$.

\section{Gauging the subsystem symmetries} \label{sec:gauging}

In this section, we examine the variety of topological phases that arise from gauging some or all of the subsystem symmetries, starting either from $|SSPT\rangle$ or $|SSET\rangle$. More precisely, starting from $|SSPT\rangle$, we can independently choose to gauge the global symmetry, lattice-plane symmetry, and dual-plane symmetry, resulting in eight possible models (including $|SSPT\rangle$). We denote the global, lattice-plane, and dual-plane symmetries as $\mathbb{Z}_2^{glob}$, $\mathbb{Z}_2^{sub_1}$, and $\mathbb{Z}_2^{sub_2}$, respectively. We use this notation with the understanding that there are a sub-extensive number of subsystem symmetry generators, so $\mathbb{Z}_2^{sub_{1/2}}$ describes the local action of the symmetry, not the total symmetry group. It is known that gauging subsystem symmetries can result in models with fracton topological order \cite{Williamson2016,Vijay2016,Shirley2019}. Here, by ``fracton'' topological order (or more succinctly fracton order), we mean any model in which \textit{all} topological excitations have restricted mobility of some sort. In our definition, this includes systems like stacks of 2D toric codes, which are usually considered to be trivial as fracton orders.
We use the usual terminology of fracton, lineon, and planon to describe point-like topological excitations which are fully immobile, constrained to a 1D line, or constrained to a 2D plane, respectively. Later, we consider models in which such excitations coexist with fully-mobile excitations, and we refer to the order in these models as ``panoptic''~\cite{Prem2019}.

Throughout this section, we do not explicitly perform the gauging as we did in the previous section (except in one case). Rather, we rely on our understanding of $|SSPT\rangle$ and its symmetries to determine both the mobility and type of symmetry enrichment of the resulting topological excitations. The key properties of the gauged models are summarized in Table~\ref{tab:table}.

\subsection{Symmetry defects and gauging} \label{sec:defects}

\begin{table*}[]
\caption{\label{tab:table}Summary of gauged models. Gauge fluxes for the global symmetry are always loop-like, and gauge fluxes for subsystem symmetries are planons, but they may be composites of excitations with lower mobility.}
\begin{tabular*}{\linewidth}{c @{\extracolsep{\fill}}c @{\extracolsep{\fill}}c @{\extracolsep{\fill}}c}
\hline \hline
Gauged symmetry & Type of order & Mobility of gauge charges & Effect of ungauged symmetries  \\ \hline 
{None} & SSPT & ---  & Mixed glob./sub. boundary anomaly \\ 
{Global} & SSET & Unrestricted mobility & Fractionalization of loops \\
{Lattice-plane}  & Fracton & {Lineon} & Mixed glob./sub. fractionalization \\
{Dual-plane}  & Fracton & {Planon} & Mixed glob./sub. fractionalization \\
Global + Lattice-plane  & Panoptic & Unrestricted + Lineon & Attaches charges to fluxes \\
Global + Dual-plane  & Panoptic & Unrestricted + Planon & Attaches charges to fluxes \\
Lattice-plane + Dual-plane  & Fracton & {Fracton} & Fracton permutation \\ 
{All}  & Panoptic & Unrestricted + Fracton &  {---} \\ \hline \hline
\end{tabular*}
\end{table*}

\begin{figure}[t]
\centering
\subfigure[]{\label{fig:defect1}\includegraphics[scale=0.135]{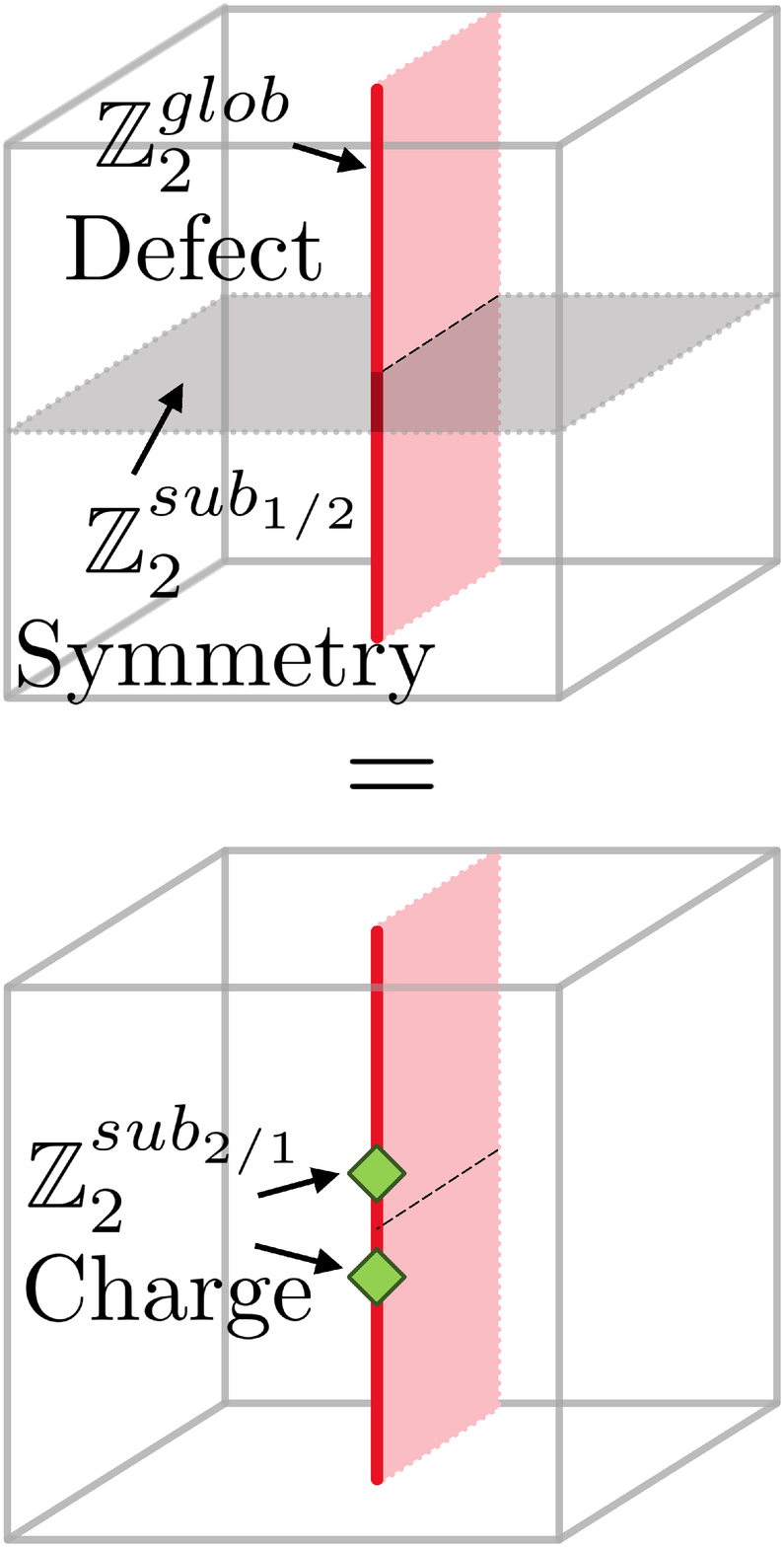}} \hfill
\subfigure[]{\label{fig:defect2}\includegraphics[scale=0.135]{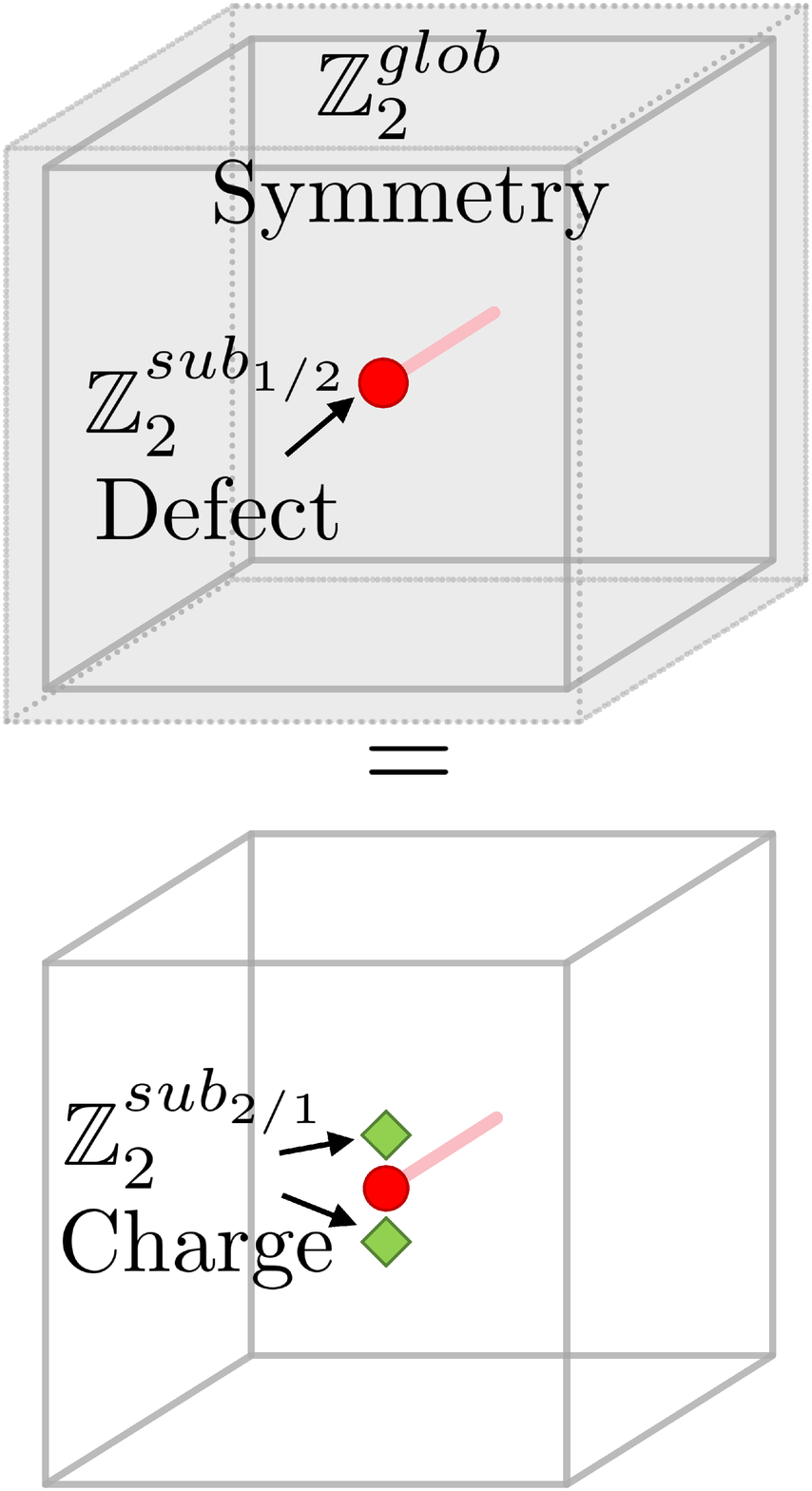}} \hfill
\subfigure[]{\label{fig:defect3}\includegraphics[scale=0.135]{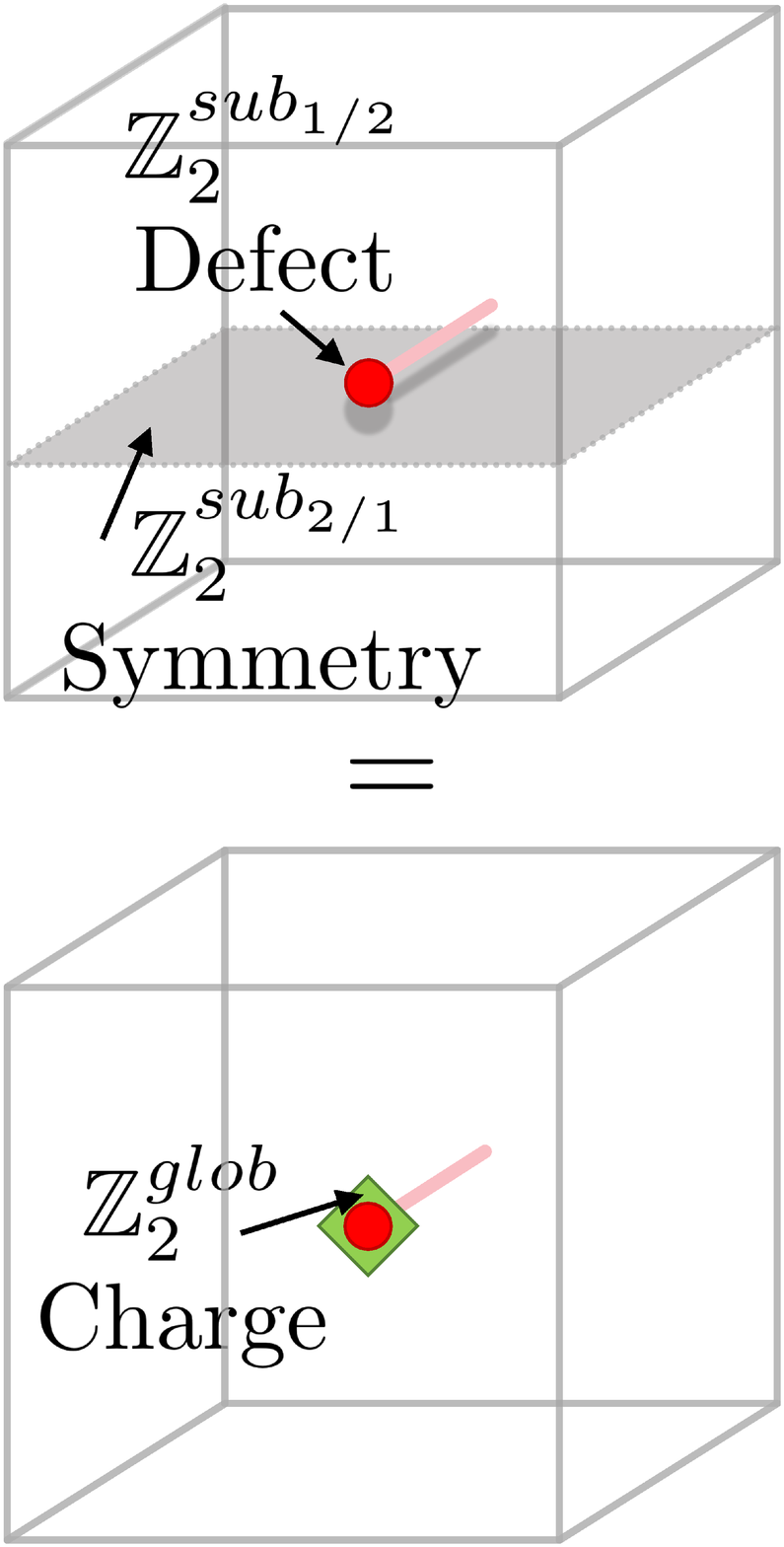}}
\caption{Interplay between symmetries, symmetry defects, and symmetry charges in $|SSPT\rangle$, as revealed by Eqs.~(\ref{eq:slant}) and (\ref{eq:subslant}). (a) Acting with a $\mathbb{Z}_2^{sub_{1/2}}$ symmetry on the line-like $\mathbb{Z}_2^{glob}$ flux creates $\mathbb{Z}_2^{sub_{2/1}}$ charges on neighbouring planes. (b) $\mathbb{Z}_2^{glob}$ symmetry on a point-like $\mathbb{Z}_2^{sub_{1/2}}$ flux creates $\mathbb{Z}_2^{sub_{2/1}}$ charges on neighbouring planes. (c) $\mathbb{Z}_2^{sub_{2/1}}$ symmetry on a neighbouring $\mathbb{Z}_2^{sub_{1/2}}$ flux attaches a $\mathbb{Z}_2^{glob}$ charge to the flux.}
\label{fig:defects}
\end{figure}

To understand the effects of gauging, we need to understand the symmetry defects of $|SSPT\rangle$.
To define a symmetry defect, we first define domain wall operators. Consider the operator $U_\mathcal{R}$ which applies some symmetry operator to a compact region $\mathcal{R}$ of the lattice. For a global symmetry, $\mathcal{R}$ is some 3D region, whereas for our subsystem symmetry, $\mathcal{R}$ would be a 2D region confined to a plane. Then, for Hamiltonians composed of local, symmetric terms, the only non-trivial effect of $U_\mathcal{R}$ will be near the boundary $\partial \mathcal{R}$ of $\mathcal{R}$. Therefore, we can write $U_\mathcal{R} \cong V_{\partial\mathcal{R}}$, where `$\cong$' indicates that the two operators act in the same way within the ground state subspace, and $V_{\partial\mathcal{R}}$ is an operator acting on $\partial\mathcal{R}$ which we call the domain wall operator. $V_{\partial\mathcal{R}}$ is precisely a membrane of $CZ$'s for $\mathbb{Z}_2^{glob}$, while for the planar subsystem symmetries it is a 1D loop operator also consisting of $CZ$'s. Symmetry defects are defined to appear at the boundaries of open domain wall operators. The $\mathbb{Z}_2^{glob}$ symmetry defects are closed 1D loops, while the $\mathbb{Z}_2^{sub_{1/2}}$ symmetry defects are point-like, see Fig.~\ref{fig:defects}.

The symmetry defects carry important information about what happens after gauging the symmetry. After gauging, domain wall operators are proliferated, and the symmetry defects become deconfined topological excitations that we call gauge fluxes. As we saw in Section \ref{sec:sset}, $\mathbb{Z}_2^{glob}$ gauge fluxes are mobile loop-like excitations. Gauge fluxes corresponding to generators of $\mathbb{Z}_2^{sub_{1/2}}$, on the other hand, are point-like and can only move within a single plane without creating additional excitations, \textit{i.e.} they are planons. This is because the $\mathbb{Z}_2^{sub_{1/2}}$ symmetry defects are themselves point-like and mobile only within a given plane. However, we find that in some cases these planons can be decomposed into a pair of excitations of lower mobility, as in the X-cube model~\cite{Vijay2016}. 
This is due to the symmetry domain walls decomposing further, such as a planar domain wall decomposing into a product of two cage-edge domain walls as in X-cube~\cite{Prem2019a}.
The properties of the symmetry defects before gauging, including the action of symmetry on them, determine the braiding and fusion statistics of the gauge fluxes, as well as possible symmetry fractionalization under any ungauged symmetries~\cite{Barkeshli2019,NewSETPaper2017,Garre-Rubio2017}.

The other type of topological excitations that emerge from gauging are called the gauge charges. The gauge charges are gauged versions of symmetry charges, which are objects that locally anti-commute with the symmetry, corresponding in our case to a single $Z$ operator on the lattice. 
There are also composites of gauge charges and existing topological excitations such as gauge fluxes, sometimes referred to as dyons.  These arise from the symmetry charges created by the action of symmetry on gauge fluxes or defects that are then promoted to gauge charges via gauging. Even for an abelian symmetry group this can lead to nonabelian topological excitations after gauging due to permutation or fractionalization \cite{Barkeshli2019}.
As described in Ref.~\cite{Shirley2019}, the mobility of gauge charges can be determined directly from the spatial structure of the subsystem symmetries. Namely, if a symmetry charge is acted on by planar symmetries in one, two, or three orthogonal directions, then the corresponding gauge charge will be a planon, lineon, or fracton, respectively. Relations between the symmetry generators that imply parity conservation for the number of subsystem charges on all planes in orthogonal directions can be used to determine whether such lineons and fractons are irreducible, meaning they are not a composite of particles of higher mobility~\cite{song2018twisted,Pai2019,Brown2019}.

The slant products calculated in Section~\ref{sec:sspt} can be used to determine the relationship between the symmetry generators, symmetry charges, and symmetry defects. Indeed, although we initially performed these calculations in Section~\ref{sec:sspt} to understand the action of the symmetry on the boundary in the presence of symmetry flux, a symmetry flux is in fact inserted using a domain wall operator, which terminates on the boundary via a symmetry defect. So we were actually calculating the action of the symmetries on the symmetry defects, regardless of whether or not these defects were pushed to the boundary. The algebra of symmetry actions on defects in $\ket{SSPT}$, as described by the slant products, allows us to calculate the stable topological excitations when a (subsystem) symmetry is gauged, by condensing domain walls and projecting topological excitations and defects onto irreps of the appropriate symmetry actions. We can further calculate the action of the remaining ungauged symmetry on the topological excitations via the algebra formed by the remaining symmetry operators and the fluxes and charges of the gauged symmetries.

More precisely, for the SSPT model, if symmetries $b$ and $c$ anticommute on a $a$ defect, \textit{i.e.} $\chi_a(b,c)/\chi_a(c,b)=-1$, then acting with $b$ ($c$) on an $a$ defect creates a $c$ ($b$) symmetry charge. This will manifest either as a fractionalization or permutation of the excitations in the gauged theory, depending on which symmetries are gauged. Specifically, if $a$ is among the gauged symmetries, while $b$ and $c$ are not, the $a$ gauge flux will carry a fractional charge under $b$ and $c$. If $a$ and $b$ are gauged and $c$ is not, then acting with $c$ symmetry on an $a$ ($b$) gauge flux attaches a $b$ ($a$) gauge charge. The action of the symmetries on the symmetry defects of $|SSPT\rangle$ (and therefore on the corresponding gauge fluxes of the gauged models) can be determined by Eqs.~(\ref{eq:slant}) and (\ref{eq:subslant}), and is summarized in Fig.~\ref{fig:defects}.

\subsection{Gauging only subsystem symmetries: Fracton~order} \label{sec:gaugingsubsystem}

We now turn to gauging the subsystem symmetries of $|SSPT\rangle$. The resulting models all have fracton topological order, characterized by topological excitations with restricted mobility. The ungauged symmetries will either permute the excitations or fractionalize on them, depending on whether we gauge all or only some of the subsystem symmetries.

\subsubsection{Gauging lattice-plane symmetries} \label{sec:frac1}

Gauging $\mathbb{Z}_2^{sub_1}$ results in a non-trivial fracton model that is not obviously equivalent to any known models. Symmetry charges before gauging are acted on by two perpendicular subsystem symmetries, so the gauge charges are lineons. These are irreducible as the product of all lattice-plane symmetries is the identity, so there is a charge parity conservation symmetry for all lattice-plane symmetries taken together. 
Interestingly, due to how the ungauged symmetries act on the gauge fluxes by creating symmetry charges (Figs.~\ref{fig:defect2},\ref{fig:defect3}), the gauge fluxes transform as a non-trivial projective representation under $\mathbb{Z}_2^{glob}$ and adjacent $\mathbb{Z}_2^{sub_2}$ symmetry generators. Therefore, this model represents a fracton model where excitations carry a fractional charge under the combination of global and subsystem symmetries.

\subsubsection{Gauging dual-plane symmetries} \label{sec:frac2}

Each edge qubit is acted on by only one generator of $\mathbb{Z}_2^{sub_2}$. Therefore, the gauge charges are planons, and furthermore we can gauge the symmetry in each dual-plane separately, such that we arrive at a stack of decoupled 2D toric codes in all three directions, one for each dual plane, which can be viewed as a rather trivial fracton model. The layers are coupled via the body qubits in such a way that the $\mathbb{Z}_2^{sub_2}$ gauge fluxes in a given plane transform as a non-trivial projective representation under the adjacent $\mathbb{Z}_2^{sub_1}$ generators and $\mathbb{Z}_2^{glob}$ (Figs.~\ref{fig:defect2},\ref{fig:defect3}).

\subsubsection{Gauging all subsystem symmetries} \label{sec:frac3}

\begin{figure}[t]
\centering
\includegraphics[width=\linewidth]{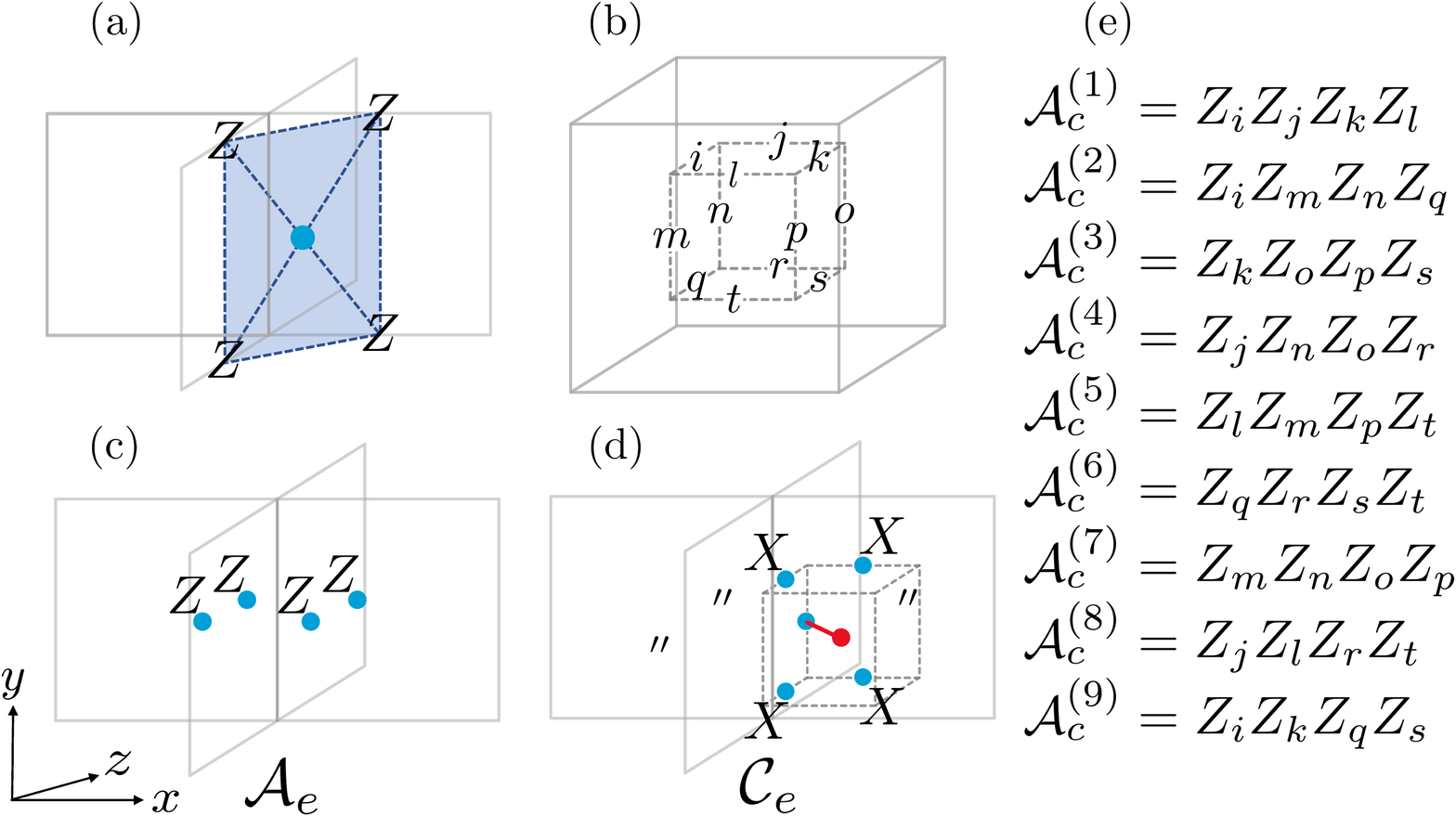}
\caption{Hamiltonian terms of the fracton model obtained from gauging subsystem symmetries of $|SSPT\rangle$. (a) The minimal interaction term symmetric under all subsystem symmetries. The gauge qubit (blue dot) is placed at the nexus of the interaction. (b) Labelling of the 12 gauge qubits within a cube, which live on the edges of a smaller inscribed cube. (c) The flux term $\mathcal{A}_e$ consists of four $Z$ operators acting on the gauge qubits closest to $e$, where $e$ is the central vertical edge.  (d) The Hamiltonian term $\mathcal{C}_e$ consists of sixteen $X$ operators acting on gauge qubits and four $CZ$ operators (denoted by a red line) that connect a gauge qubit to a body center qubit. The interaction is only shown in one quadrant for clarity; it acts in an analogous (rotated) manner in all 4 cubes surrounding the edge $e$ as indicated by the dashes. (e) The nine flux terms within a cube. $Z_{\bullet}$ is shorthand for $Z_{(c,\bullet)}$. The flux terms $\mathcal{A}_e$ and $\mathcal{A}_c^{(i)}$ satisfy a number of relations which determine the mobility of gauge fluxes, such as the product of all $\mathcal{A}_e$ and $\mathcal{A}_c^{(7)}$ in an $xz$ dual-plane; $\mathcal{A}_e$, $\mathcal{A}_c^{(4)}$, $\mathcal{A}_c^{(5)}$ in a $yx$ lattice-plane; and $\mathcal{A}_e$, $\mathcal{A}_c^{(2)}$, $\mathcal{A}_c^{(3)}$ in a $yz$ lattice plane.}
\label{fig:fractonham}
\end{figure}

Gauging all subsystem symmetries of $|SSPT\rangle$ results in a model with fracton excitations, where the global symmetry permutes the fractons. This permutation action can be understood by the fact that the global symmetry, when acting on a lattice-plane (dual-plane) symmetry defect, attaches a pair of symmetry charges to the neighbouring dual-planes (lattice-planes). Hence, in the gauged model, the global symmetry attaches gauge charges to gauge fluxes (Fig.~\ref{fig:defect2}). As in the general case, the gauge flux is a planon. However, in this case, a planar domain wall can be decomposed into a pair of domain walls and thus the planon can be split into a pair of fractons. We can isolate these two fractons by considering the symmetry defect corresponding to a stack of dual-plane symmetries. Gauging this defect gives two fractons, one at the top of the stack and one at the bottom. The fact that these are fractons follows from relations between the gauge constraints, as we elaborate on in Fig.~\ref{fig:fractonham}. Applying the global symmetry attaches a gauge charge to each fracton. Since the symmetry charges are acted on by planar symmetries in all three directions, the gauge charges are fractons. In particular, there are relations between all subsystem symmetries in two orthogonal directions that imply the fractons are irreducible. Therefore, the global symmetry action causes fracton permutation.

This permutation action will be useful for understanding the model obtained by gauging all symmetries in Section \ref{sec:pan3}, so here we will explicitly write down the gauged Hamiltonian.
To perform this gauging, we follow the procedure for gauging subsystem symmetries outlined in Ref.~\cite{Shirley2019}. First, we identify the minimal coupling terms which commute with all subsystem symmetries, which in the present case takes the form of four-body interactions near an edge, pictured in Fig.~\ref{fig:fractonham}. The gauge qubits are placed in the nexus of these interactions as shown in Fig.~\ref{fig:fractonham}. We therefore have four qubits for each edge $e$, one for each cube $c$ surrounding $e$. We may therefore label gauge qubits uniquely by a pair $(c,e)$. We can view the gauge qubits as living on the edges of a smaller cube within each cube of the lattice, and we use the labelling in Fig.~\ref{fig:fractonham} to reference each of the twelve gauge qubits in a given cube. Following Ref.~\cite{Shirley2019}, we now express existing Hamiltonian terms $\widetilde{B}_c$ and $\widetilde{C}_e$ in terms of the new gauge qubits, resulting in the terms $\mathcal{B}_c$ and $\mathcal{C}_e$. Furthermore, we add additional flux terms which enforce a zero-flux constraint on the gauge qubits, as determined by relations between the minimal coupling terms \cite{Shirley2019}. There is one flux term $\mathcal{A}_e$ for each edge, and nine flux terms $\mathcal{A}^{(k)}_c$ for each cube $c$. We define $\mathcal{C}_e$ and the flux terms in Fig.~\ref{fig:fractonham}. The precise form of $\mathcal{B}_c$ is not required for our purposes here, as it is not associated with any topological excitations. Overall, the gauged Hamiltonian is,
\begin{align} \label{eq:hfrac}
H_{frac} = -\sum_{c\in C} \mathcal{B}_c - \sum_{e\in E} \mathcal{C}_e - \sum_{e\in E} \mathcal{A}_e - \sum_{c\in C} \sum_{k=1}^9 \mathcal{A}^{(k)}_c\ .
\end{align}

Now let us investigate the excitations of $H_{frac}$. Due to the sheer number of terms, and the fact that they are not all independent, the full spectrum of excitations is tedious to describe. Instead, let us focus on two types which allow us to see the non-trivial action of the global symmetry. Gauge charges, \textit{i.e} violations of $\mathcal{C}_e$,  can be created at the corners of a membrane operator,
\begin{equation}
\mathcal{S}^\mathrm{e}_\mathcal{R} = \prod_{c\in \mathcal{R}} Z_{(c,j)} Z_{(c,l)}\ ,
\end{equation}
where $\mathcal{R}\subset C$ is a rectangle of cubic cells in the $xz$ plane, and excitations appear on the edges at the four corners of $\mathcal{R}$. Gauge fluxes corresponding to violations of $\mathcal{A}_e$ can be created by the following membrane operator,
\begin{equation}
\mathcal{S}^{\mathrm{m}}_\mathcal{R} = \prod_{c\in \mathcal{R}} X_{(c,m)}X_{(c,n)}X_{(c,o)}X_{(c,p)} CZ_{c,(c,j)} CZ_{c,(c,l)},
\end{equation}
where gauge fluxes again appear at the four corners of $\mathcal{R}$. The fact that these are fractons follows from the three orthogonal relations among flux terms described in Fig.~\ref{fig:fractonham}. Now we observe that the global symmetry $X_C$ permutes the membrane operators,
\begin{equation} \label{eq:fracperm}
X_C \mathcal{S}^{\mathrm{m}}_\mathcal{R} X_C^\dagger = \mathcal{S}^{\mathrm{m}}_\mathcal{R} \mathcal{S}^\mathrm{e}_\mathcal{R}\ .
\end{equation}
This shows that acting with the global symmetry attaches gauge charges onto gauge fluxes, thereby permuting the fractons of the gauged theory. We remark that Eq.~\ref{eq:fracperm} is implied directly by a similar equation describing the action of the global symmetry on the symmetry defect of a stack of dual-planes, and that similar equations hold for all permutation actions described in the following section.

\subsection{Gauging global and subsystem symmetries: Panoptic~order} \label{sec:gaugingboth}

In this section, we consider gauging the global symmetry of $|SSPT\rangle$ along with some or all of the subsystem symmetries. Equivalently, we are gauging the subsystem symmetries of $|SSET\rangle$. In each case, we will find that fully mobile point-like and loop-like excitations, coming from the gauged global symmetry, coexist with the restricted-mobility excitations coming from the gauged subsystem symmetries. Such a system was dubbed to have ``panoptic'' order in Ref.~\cite{Prem2019}.

\subsubsection{Gauging lattice-plane and global symmetries} \label{sec:pan1}
 
If we gauge $\mathbb{Z}_2^{sub_1}$ and $\mathbb{Z}_2^{glob}$, we end up with a model that is, in the absence of symmetry, equivalent to a stack of the 3D toric code and the fracton model from Section \ref{sec:frac1}. The ungauged $\mathbb{Z}_2^{sub_2}$ symmetry couples the two models by permuting excitations between them. Specifically, a $\mathbb{Z}_2^{sub_2}$ generator creates  $\mathbb{Z}_2^{sub_1}$ gauge charges on a pair of planes wherever it intersects the loop-like $\mathbb{Z}_2^{glob}$ gauge fluxes (Fig.~\ref{fig:defect1}), and the same generator attaches a $\mathbb{Z}_2^{glob}$ gauge charge to $\mathbb{Z}_2^{sub_1}$ gauge fluxes in adjacent planes (Fig.~\ref{fig:defect3}). Thus, the symmetry enrichment manifests as an interesting permutation involving fully mobile excitations and those of restricted mobility.

\subsubsection{Gauging dual-plane and global symmetries} \label{sec:pan2}

Gauging $\mathbb{Z}_2^{sub_1}$ and $\mathbb{Z}_2^{glob}$ gives a model in the same topological phase as a stack of the 3D toric code and layers of 2D toric codes in all three directions. Similar to the previous model, the remaining $\mathbb{Z}_2^{sub_1}$ symmetry generators permute excitations between the 2D and 3D toric codes by attaching $\mathbb{Z}_2^{sub_2}$ gauge charges to the loop-like $\mathbb{Z}_2^{glob}$ gauge fluxes (Fig.~\ref{fig:defect1}) and  by attaching $\mathbb{Z}_2^{glob}$ gauge charges to $\mathbb{Z}_2^{sub_1}$ gauge fluxes in adjacent planes (Fig.~\ref{fig:defect3}).

\subsubsection{Gauging all symmetries} \label{sec:pan3}

To understand the model obtained by gauging all symmetries, it is easiest to start from $H_{frac}$ in Eq.~\ref{eq:hfrac}, where all subsystem symmetries have been gauged. Then, what remains is to gauge the global symmetry, which we saw enacts a non-trivial permutation on the fractons of the model. For this, we can use the general arguments of Ref.~\cite{Prem2019} (see also Ref.~\cite{Bulmash2019}). Therein, it is argued that gauging a fracton permuting symmetry results in a model with non-abelian fractons. Furthermore, gauge fluxes will be loop-like excitations that braid non-trivially with the excitations of reduced mobility. Since we expect the gauging of different symmetries to commute (as is indeed the case in Appendix \ref{app:2d}), we can conclude that gauging the subsystem symmetries of the SSET will result in the same panoptic order with non-abelian fractons.

It is interesting to compare to the model obtained by gauging the layer-swap symmetry of the bilayer X-Cube model, as presented in Refs.~\cite{Prem2019,Bulmash2019}, which also has panoptic order with non-abelian fractons. A potential equivalence between this model and our own is suggested by the equivalence between the model obtained by gauging the layer-swap symmetry of the bilayer 2D toric code \cite{Prem2019,Bulmash2019}, and the model obtained by gauging all symmetries of the non-trivial $\mathbb{Z}_2\times\mathbb{Z}_2\times\mathbb{Z}_2$ SPT with type-III cocycle~\cite{Propitius1995} (which is somewhat analogous to $|SSPT\rangle$), as discussed in Appendix~\ref{app:2d}. Therefore it is tempting to conjecture that the gauged bilayer X-Cube model and the model obtained by gauging all symmetries of $|SSPT\rangle$ are equivalent as gapped phases of matter.

\section{Discussion \& Conclusions} \label{sec:conclusion}

We investigated the phenomenon of subsystem symmetry enrichment in 3D systems. We began with a base model possessing SPT order under a mix of global and planar subsystem symmetries. By gauging the global symmetries of this model, we obtained a topological model with loop-like excitations that carry fractional charge of the subsystem symmetries, which we called an example of SSET order. We showed that this fractionalization leads to a extensive degeneracy of the excitations, as well as an increased value of the topological entanglement entropy. We then considered also gauging the subsystem symmetries of the base model, resulting in a network of models all related by gauging and ungauging symmetries (Fig.~\ref{fig:flow}). Using the algebra of the symmetry defects of the SPT model, we were able to understand the nature of each gauged model, uncovering several distinct types of subsystem symmetry enrichment. In particular, we found models supporting mixed global and subsystem symmetry fractionalization, permutation between mobile and restricted mobility excitations, and a model with non-abelian fractons. 

To conclude, we present the first steps towards a general theory of subsystem symmetry enrichment, which allow us to argue against the existence of nontrivial subsystem symmetry enrichment in 2D systems. We then give an outlook on generalizations and possible applications of our results.

\subsection{Towards a general theory of subsystem symmetry enrichment}

As discussed briefly in the introduction, symmetry enrichment is defined by a symmetry action on topological excitations and defects, which can involve permutation and fractionalization. For subsystem symmetries, the same is true. However, as demonstrated in our examples, there is an additional non-trivial interplay between the spatial structure of the subsystem symmetries and the mobility and geometry of the topological excitations.
The mobility restrictions on fractons are often formulated in terms of abelian conservation rules for the particles supported on various subsystems, which may be planar or fractal~\cite{Pai2019,Brown2019}. 
Furthermore, such conservation rules on deformable co-dimension $k$ subsystems encode that a local excitation must appear as part of an extended $k$-dimensional excitation. 
The spatial intersection of these subsystem conservation rules and the subsystem symmetry generators lead to constraints on possible consistent symmetry actions that generalize those of a global symmetry in 2D. 

To capture this we introduce the concept of fusion rules restricted to subsystems, assuming no fusion degeneracy for simplicity, we write 
$N_{a b}^{c} |_{S} = 0,1,$ 
to indicate whether topological excitations $a$ and $b$, supported on a subsystem $S$, may fuse to $c$, also supported on $S$, creating and annihilating no further excitations on $S$. We remark that this definition allows arbitrary excitations to be created and annihilated outside $S$, and applies to segments of extended topological excitations that are supported on $S$. For example, a looplike excitation restricts to pointlike excitations where it intersects a codimension-1 subsystem symmetry. 
This allows us to generalize symmetry enrichment with global symmetries to include subsystem symmetries, by replacing constraints on the action of a global symmetry from the quasiparticle fusion rules with constraints on the action of subsystem symmetries from the fusion rules restricted to the appropriate subsystems. 

This paints a general picture for how subsystem symmetries can act on topological phases, via permutation and generalized projective representations that satisfy the consistency equations coming from all restricted fusion rules, or equivalently conservation laws. This generalizes the familiar classification of 2D SET phases. 
The projective representation of the full subsystem symmetry group should further respect locality in the sense that the action of subsystem generators that are far separated in space should commute, while those that act on a common excitation in the same spatial region need not commute. 

As an example consider loop-like excitations in 3D, which obey the 1-form conservation law that every sphere is pierced by a loop an even number of times. Codimension-1 symmetries may act via nontrivial projective representations with $\mathbb{Z}_2$ fusion (as in Section \ref{sec:sset}) and locally permute the string excitations, which could involve attaching particles (as in Section \ref{sec:pan1}) or a segment of a general 1-dimensional defect~\cite{PhysRevB.91.245131,else2017cheshire} with $\mathbb{Z}_2$ fusion rules. 
For a second example, fractons with planar charge conservation rules such as in the X-cube model~\cite{PhysRevB.81.184303,Vijay2015} and other foliated fracton phases~\cite{Shirley2018,Slagle2018} can be acted upon nontrivially by planar subsystem symmetries that are aligned with the planar conservation rules \cite{You2018a}.

\subsubsection{No nontrivial SSET in 2D}

Our discussion of SSETs above leads to the conclusion that there can be no nontrivial intrinsic SSETs in 2D, in other words all such SSETs are equivalent to a stack of a conventional SET and an SSPT. 
To see this we first point out that there are no fracton topological orders in 2D and hence all topological particles are fully mobile~\cite{Aasen2020}. 
Thus any topological charge can be moved to the complement region not acted upon by any subsystem symmetry generator. 
This implies the action of this subsystem symmetry on any topological sector must be trivial, involving no permutation nor projective representation. 
Although the string operators for the anyons may have unavoidable intersections with the subsystem symmetries, the defects created by the subsystem symmetries are in the trivial topological superselection sector, by definition of an SSET. Hence there can be no statistical processes between the anyonic string operators and subsystem symmetries either. 

The argument above immediately generalizes to rule out nontrivial subsystem symmetry-enrichment on fully mobile point charges, and nontrivial actions by codimension-$k$, or above, subsystem symmetries on fully mobile $(k-1)$-dimensional, or lower, extended excitations.

\subsection{Outlook}

Our SSET model can be obtained by decorating the 3D toric code with 2D cluster states. We can straightforwardly generalize our model by changing both the underlying 3D topological model, as well as the 2D SSPT model used to decorate. This raises the question of classification for SSET phases in 3D, and whether all phases can be captured by such decorated constructions. We remark that there should be some issues of compatibility, in that only certain combinations of topological model and SSPT are allowed. For example, in 2D SET models, the possible kinds of fractionalization that a point-like excitation can carry are restricted by its braiding statistics with other excitations~\cite{Barkeshli2019}. In analogy, we suspect that there is a connection between subsystem symmetry fractionalization in loop-like excitations and their braiding with point-like excitations. As a further generalization, it is also plausible that, by decorating with a 2D SSPT possessing 2D fractal subsystem symmetries \cite{Williamson2016,Kubica2018, Devakul2019,Stephen2019a}, one could obtain a 3D SSET enriched by 3D fractal subsystem symmetries. 
These questions are closely related to the classification of SPTs with mixed global and subsystem symmetry, where the approach of Ref.~\cite{Tantivasadakarn2019} should be applicable.  

The models analyzed in this work inherit their properties from the non-trivial interplay between global and subsystem symmetries. In particular, gauging all symmetries results in panoptic order, rather than ``pure'' fracton order in which all excitations have mobility constraints. It would be interesting to remove the global symmetries from the equation and consider systems with subsystem symmetries alone. Aside from providing novel mechanisms for subsystem symmetry fractionalization, this provides a route towards a systematic construction of ``pure'' fracton models with non-abelian fracton excitations~\cite{VijayFu2017,song2018twisted,GaugedLayer}, which would arise from gauging a symmetry-enriched fracton model in which fracton excitations carry a fractional subsystem symmetry charge. Additionally, by considering fractal subsystem symmetries, it may be possible to obtain type-II fracton models with non-abelian fractons which, as of yet, have proved elusive, previous attempts having resulted in panoptic models~\cite{Prem2019,Bulmash2019}.

Finally, we address the possible applications of our models to the storage and processing of quantum information. In general, it would be interesting to investigate whether the introduction of subsystem symmetries and subsystem symmetry defects into topological models can augment their computational capabilities, as is the case of global symmetry defects in 2D topological systems~\cite{Bombin2010,Barkeshli2013,Delaney2020}. Regarding the SSET model from Section~\ref{sec:sset}, the fact that it combines the universal (measurement-based) quantum computational power of the SSPT order \cite{Raussendorf2019,Devakul2018a,Stephen2019a,Daniel2019} together with the information storage capabilities of the topological order \cite{Kitaev2003,Dennis2002,Kubica2018a} suggests some potential applications. A relevant phenomenon that may be of use is the  emergence of a symmetry protected degeneracy on the loop-like excitations.

\vspace{.5cm}

\section*{Acknowledgements}
DTS is grateful to Norbert Schuch for interesting discussions. AD thanks Meng Cheng for useful discussions. DTS, AD, and JGR thank the Centro de Ciencias de Benasque Pedro Pascual for their hospitality during the 2019 Quantum Information workshop, where this work was initiated. This work has received funding from the European Research Council (ERC) under the European Union’s Horizon 2020 research and innovation
programme (grant nos. 636201 and 648913). DTS was supported by a fellowship from the Natural Sciences and Engineering Research Council of Canada (NSERC), and by the Deutsche Forschungsgemeinschaft (DFG) under Germany’s Excellence Strategy (EXC-2111 – 390814868). JGR acknowledges financial support from MINECO (grant MTM2017-88385-P), Comunidad de Madrid (grant QUITEMAD-CM, ref. P2018/TCS­4342), and ICMAT Severo Ochoa project SEV-2015-0554 (MINECO). 
DW acknowledges support from the Simons Foundation.

\bibliography{sset_biblio}

\newpage

\appendix

\section{Gauging and symmetry enrichment in 2D} \label{app:2d}

In this section, we discuss a model of 2D SPT order which, upon gauging different subgroups of its symmetry, displays different kinds of symmetry enrichment. This serves as a simple analogue to the 3D model discussed in the main text.

\begin{figure}[t]
\centering
\subfigure[]{\label{fig:unionjack}\includegraphics[scale=0.06]{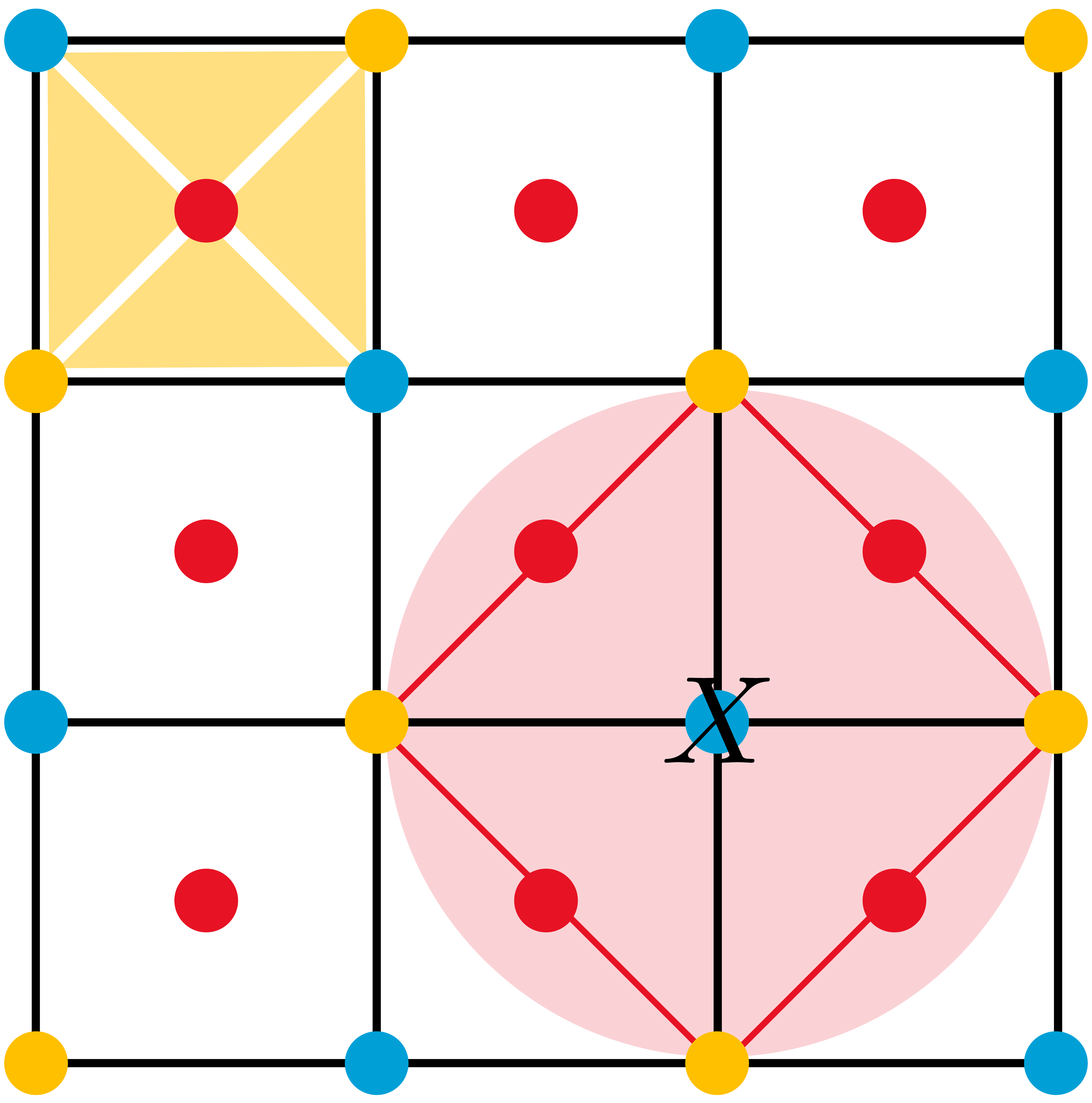}} \hfill
\subfigure[]{\raisebox{2mm}{\label{fig:ddw}\includegraphics[scale=0.08]{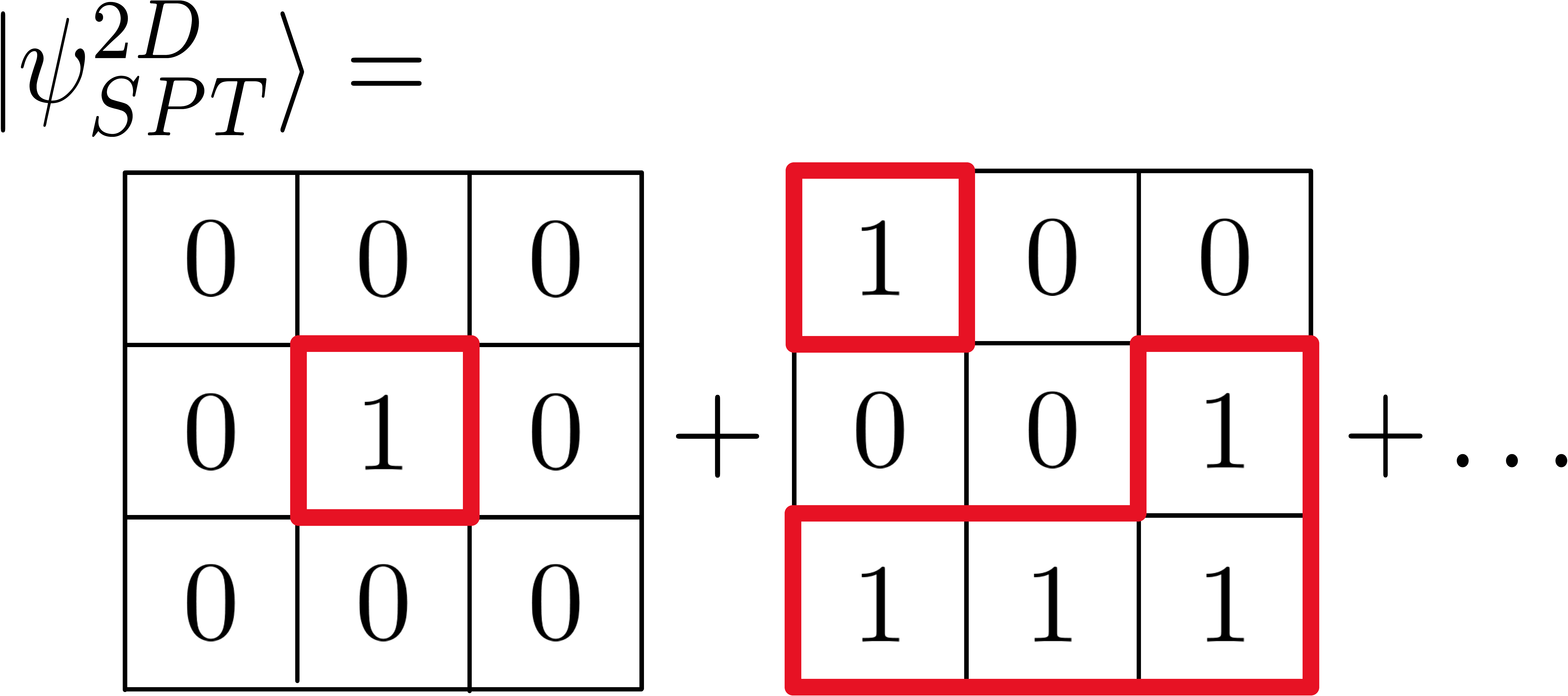}}}
\caption{(a) The Union-Jack lattice. Filled triangles represent the four $CCZ$ unitaries associated to one face. The shaded circle contains the Hamiltonian term $\widetilde{C}_v$, where $X$ acts on $v$ and red lines represent $CZ$ unitaries. (b) The decorated domain wall structure of $|SPT\rangle$. Vertex qubits are initialized in the $|+\rangle$ state, then $CZ$ unitaries are applied along thick red lines.}
\end{figure}

The starting point is the ``Union-Jack'' model of 2D SPT order~\cite{Miller2016,Yoshida2016}, which is closely related to the model of Ref.~\cite{Levin2012}. The model can be described by a simple cubic lattice with qubits living on the vertices ($V$) and faces ($F$) of the lattice, as pictured in Fig.~\ref{fig:unionjack}. We then define the state,
\begin{equation}
|SPT\rangle=\prod_{\triangle} CCZ_\triangle \, |+\rangle^{\otimes |V\oplus F|} 
\end{equation}
where the product runs over all triangles formed by triples consisting of a face qubit and two neighbouring edge qubits, as depicted in Fig.~\ref{fig:unionjack}.
We can obtain a Hamiltonian for which $|SPT\rangle$ is the unique ground state starting with trivial Hamiltonian whose ground state is $|+\rangle^{\otimes |V\oplus F|}$, 
\begin{equation}
H_{triv}=-\sum_{v\in V} X_v-\sum_{f\in F} X_f,
\end{equation}
and then conjugating it by the entangling circuit $\prod_{\triangle} CCZ_\triangle$ to obtain,
\begin{equation} \label{eq:h2dspt}
H_{SPT}=-\sum_{f\in F} \widetilde{B}_f - \sum_{v\in V} \widetilde{C}_v 
\end{equation}
where $\widetilde{C}_v$ is pictured in Fig.~\ref{fig:unionjack} and, 
\begin{equation}
\widetilde{B}_f = X_f\prod_{e\in f} CZ_{\partial e}
\end{equation}
where $CZ_{\partial e}$ acts on the two vertices touching $e$.

$|SPT\rangle$ has a global $\mathbb{Z}_2\times\mathbb{Z}_2\times\mathbb{Z}_2$ symmetry generated by the operators $X_{B}$, $X_{Y}$ and $X_F$, which apply $X$ on all even (blue) vertices, odd (yellow) vertices, and faces, respectively. The fact that $|SPT\rangle$ has non-trivial SPT order under these symmetries can be understood using a decorated domain wall (DDW) picture \cite{Chen2014}. Namely, we observe that the state can be expressed as an equal superposition over all configurations of the face qubits, with domain walls decorated by 1D cluster states, as pictured in Fig.~\ref{fig:ddw}. The 1D cluster state is an example of a state with 1D SPT order, so $|SPT\rangle$ is described by decorating the domain walls of a trivial state with lower-dimensional SPT order, which is known to produce non-trivial SPT order \cite{Chen2014}. 

Now, we look at how gauging subgroups of the total symmetry group lead to models with SET order. In one case, we find that the anyons of the gauged theory carry fractional charge of the residual symmetry, while in the other case the anyons are permuted by the symmetry. We then show that the gauging processes commute: if all symmetries are gauged, the resulting model does not depend on the order in which they are gauged.

\subsection{Gauging face symmetries}

First, let us gauge the $\mathbb{Z}_2$ subgroup generated by $X_F$, leaving behind a residual $\mathbb{Z}_2\times\mathbb{Z}_2$ symmetry group. We will show that the resulting topological order displays symmetry fractionalization. The gauging procedure maps qubits on the faces to new qubits on the edges of the lattice, such that the edge qubits take the state $|1\rangle$ on domain walls of the face qubits, and $|0\rangle$ away from them. The gauged Hamiltonian becomes,
\begin{equation} \label{eq:h2dset}
H_{SET}=-\sum_{v\in V} A_v -\sum_{f\in F} B_f -\sum_{v\in V} C_v\frac{1+A_v}{2}
\end{equation}
where,
\begin{equation}
A_v=\prod_{e\ni v} Z_e \, ,
\end{equation}
\begin{equation}
B_f=\prod_{e\in f} X_e CZ_{\partial e} \, ,
\end{equation}
\begin{equation}
C_v=X_v \prod_{v'\in n(v)} CZ_{v',e(v',v)} \, .
\end{equation}
Where $n(v)$ is the set of nearest-neighbouring vertices to $v$, and $e(v,v')$ is the edge connecting vertices $v$ and $v'$. We have projected $C_v$ onto the zero-flux subspace to ensure that $H_{SET}$ commutes with the residual $\mathbb{Z}_2\times\mathbb{Z}_2$ symmetry. This Hamiltonian can be disentangled by acting with $CCZ$ on every edge and its two vertices, resulting in a toric code Hamiltonian on the edges and a trivial Hamiltonian on the vertices. However, this circuit does not respect the $\mathbb{Z}_2\times\mathbb{Z}_2$ symmetry, and we now show that $H_{SET}$ is in fact in a different SET phase than the toric code Hamiltonian in the presence of this symmetry. We note that this same model appeared in Ref.~\cite{Ben-Zion2016}, although it was derived from a different perspective.

We recall that the ground states of the toric code can be described as loop condensates, which are equal weight superpositions of configurations in which the edge qubits in state $|1\rangle$ form closed loops. Furthermore, a configuration including an open string of 1's will create anyonic excitations, corresponding to violations of $A_v$, at the endpoints. After placing $CCZ_{e,\partial(e)}$ on every edge, these loops become decorated by 1D cluster states. Therefore, the anyons at the endpoints of open strings are accompanied by the edges of 1D SPTs. Since these edges transform non-trivially under the symmetry, the anyons fractionalize. 

More precisely, consider the following string operator,
\begin{equation}
S_\Gamma=\prod_{e\in\Gamma} X_e CZ_{\partial{e}}
\end{equation}
where $CZ_{\partial e}$ acts on the two vertices touching $e$, and $\Gamma$ is some open string of edges with terminal vertices $v_i$ and $v_f$, which we assume to both be yellow without loss of generality. If we apply this operator to a ground state $|SET\rangle$, we get an excited state with excitations  at $v_i$ and $v_f$, corresponding to $A_v=-1$. Because $A_v=-1$ at these points, the projection $\left(\frac{1+A_v}{2}\right)$ annihilates the Hamiltonian terms involving $C_{v_{i/f}}$, so we can dress the endpoints of $S_\Gamma$ with $Z$'s without changing the energy of the resulting excitations. This means we have four different string operators:
\begin{equation} \label{eq:sgamma}
S_\Gamma(a,b)= S_\Gamma Z_{v_i}^a Z_{v_f}^b,\quad a,b=0,1.
\end{equation}
Thus each anyon carries a two-fold degeneracy. Furthermore, $\mathbb{Z}_2\times\mathbb{Z}_2$ symmetry acts projectively on each anyon. To see this, consider the subspace of degenerate states $|a,b\rangle :=S_\Gamma(a,b)|SET\rangle$. We compute,
\begin{align}
&X_{B}|a,b\rangle=|a\oplus 1,b\oplus 1\rangle \nonumber \\
&X_{Y}|a,b\rangle=(-1)^a(-1)^b|a,b\rangle
\end{align}

Therefore, $X_{B}\sim X\otimes X$ and $X_{Y}\sim Z\otimes Z$ in this subspace. Since $X$ and $Z$ anticommute, each anyon carries a projective representation of $\mathbb{Z}_2\times\mathbb{Z}_2$, which demonstrates the fractionalization.

\subsection{Gauging vertex symmetries}

Now, we instead gauge the $\mathbb{Z}_2\times \mathbb{Z}_2$ symmetry subgroup generated by $X_{B}$ and $X_{Y}$. For convenience, denote the subsets of all blue/yellow vertices by $V_{B/Y}\subset V$. Notice that $V_{B/Y}$ each form a rotated square lattice. We can therefore gauge each $\mathbb{Z}_2$ factor individually in the same way that we gauged $X_F$ in the previous subsection, such that the vertex qubits are mapped onto a pair of new qubits on each face. This results in three qubits per face which we label $f_{r,b,y}$ with $f_r$ labelling the ungauged face qubit.

The gauged Hamiltonian reads,
\begin{align}
H'_{SET}=&-\sum_{v\in V_B} \left(A^y_v -B^b_v\frac{1+A^y_v}{2}\right) \nonumber \\ 
&- \sum_{v\in V_Y} \left(A^b_v -B^y_v \frac{1+A^b_v}{2}\right)  -\sum_{f\in F} C'_f ,
\end{align}
where, if we let $F_v\subset F$ denote the four faces surrounding vertex $v$,
\begin{equation}
A^{b/y}_v=\prod_{f\in F_v} Z_{f_{b/y}} \ ,
\end{equation}
\begin{equation}
B^{b/y}_v=\prod_{f\in F_v} X_{f_{b/y}} CZ_{f_r,f_{y/b}} \ ,
\end{equation}
\begin{equation}
C'_f=X_{f_r} CZ_{f_b,f_y} \ .
\end{equation}
We have again projected certain terms onto the zero-flux subspace to ensure the Hamiltonian is symmetric. If we apply $CCZ_{f_r,f_b,f_y}$ to every face, this Hamiltonian is disentangled to two copies of the toric code on the blue and yellow face qubits, and a trivial Hamiltonian on the red face qubits. Once again, this unitary does not respect the residual $\mathbb{Z}_2$ symmetry. 

This time, the residual symmetry acts in a way that permutes the anyons of the two toric codes, rather than fractionalizing. To see this, consider the following string operator,
\begin{equation}
Q_\Lambda=\prod_{f\in \Lambda} X_{f_b} CZ_{f_r,f_y} ,
\end{equation}
where $\Lambda$ is a connected path of faces. If we apply this operator to a ground state $|\psi_{g}\rangle$ of $H'_{SET}$, we get an excitated state with a pair of excitations of $A^b_v$ at the two yellow vertices at the endpoints of the path. Now, if we apply the symmetry to this state, we find,
\begin{equation}
X_F Q_\Lambda |\psi_g\rangle = Q_\Lambda \left(\prod_{f\in\Lambda} Z_{f_y}\right)  X_F |\psi_g\rangle = Q_\Lambda  \prod_{f\in\Lambda} Z_{f_y} |\psi_g\rangle
\end{equation}
The string operator $\prod_{f\in\Lambda} Z_{f_y}$ acting on $|\psi_g\rangle$ creates excitations of $B^y_v$ at the same yellow vertices. Therefore, acting with $X_F$ permutes the anyons of the gauge theory. We can repeat the same procedure starting with different string operators. If we let $\mathrm{m}_{b/y}$ and $\mathrm{e}_{b/y}$ denote the anyons associated to violations of $A^{b/y}_v$ and $B^{b/y}_v$, respectively, then the residual $\mathbb{Z}_2$ symmetry permutes anyons by attaching to every $\mathrm{m}$ particle an $\mathrm{e}$ particle of opposite color, \textit{i.e.} $\mathrm{m}_{b/y}\leftrightarrow \mathrm{m}_{b/y}\mathrm{e}_{y/b}$.

\subsection{Gauging all symmetries}

We can also consider gauging the whole $\mathbb{Z}_2\times\mathbb{Z}_2\times\mathbb{Z}_2$ symmetry group. While we will not perform this calculation explicitly, we can use some general results to determine the resulting gauge theory. First, suppose we start from $H_{SET}$ and gauge the fractionalized $\mathbb{Z}_2\times\mathbb{Z}_2$ symmetry. In general, when a symmetry acts on the anyons as a higher-dimensional representation, gauging it results in non-abelian anyons \cite{Barkeshli2019}. The fractionalization pattern observed for $H_{SET}$ corresponds precisely to that shown in Fig. 2(c) of Ref.~\cite{Garre-Rubio2019}, and it is shown therein that gauging the symmetry results in a model with $D_8$ topological order, where $D_8$ is the symmetry group of a square. 

Now, let us rather start from $H'_{SET}$ and gauge the anyon-permuting symmetry. In Ref.~\cite{Prem2019}, it was shown in general that gauging an anyon-permuting symmetry leads to non-abelian anyons. The symmetry considered in Ref.~\cite{Prem2019} was a layer swap of a bilayer toric code, which appears superficially different from the permutation observed for $H'_{SET}$. However, upon a relabelling of anyons that preserves the braiding and fusion rules, the two symmetries turn out to be the same. Therefore, the resulting gauge theory should be the same as the one found in Ref.~\cite{Prem2019}, namely $D_8$ topological order.

As the two above cases show, we expect a non-abelian $D_8$ topological order after gauging all symmetries of $|SPT\rangle$, regardless of the order in which the symmetries are gauged. This reinforces the intuition that the gauging operations for a pair of commuting symmetry subgroups should commute. 

\subsection{Symmetry defect analysis}

Now, we show that examining symmetry defects in the ungauged model allows us to draw the same conclusions without explicitly gauging the symmetries. Using the definition of symmetry defects from Section \ref{sec:defects}, we find that the symmetry defects associated to $X_F$ appear at the ends of 1D lines consisting of $CZ$'s on a path along the edges of the lattice. Then, we find that the endpoint of the $X_F$ defect line transforms projectively under the $X_{B/Y}$ symmetries, in the exact same way as the string operator $S_\Gamma$ in Eq.~\eqref{eq:sgamma}. The defects associated to $X_{B}$ ($X_{Y}$) appear at the ends of strings of $CZ$s on a path alternating between face qubits and yellow (blue) vertex qubits. Consider the $X_B$ defects, with endpoints on yellow vertex qubits. Then, the $X_F$ symmetry transforms the defects by dressing these endpoints with $Z$ operators. These correspond to $X_Y$ symmetry charges. If we were to then gauge the $\mathbb{Z}_2\times\mathbb{Z}_2$ symmetry generated by $X_{B/Y}$, these two charges would become a pair of anyons. Therefore, the permutation action of the symmetry in the gauged theory can be seen by how the symmetry decorates the symmetry defects with symmetry charges.

\section{Boundary Hamiltonians for $|SSPT\rangle$} \label{app:boundary}

In this Appendix, we consider some possible Hamiltonians which respect the boundary symmetries of $|SSPT\rangle$. Recall that, on the boundary, the planar subsystem symmetries act like lines of $X$ operators, while the global symmetry acts like $CZ$'s between neighbouring edges. The two simplest Hamiltonians that respect these symmetries are,
\begin{align}
&H_{cSSPT}=-\sum_{e\in \partial_E} X_e-\sum_{e\in \partial_E} X_e\prod_{e'\in n(e)} Z_{e'} \\
&H_{SSB}=-\sum_{e\in \partial_E} \prod_{e'\in n(e)} Z_{e'}
\end{align}
where $\partial_E$ denotes the set of edges on the boundary, and $n(e)$ contains the set of four edges that are nearest-neighbouring edges to $e$, as measured by distance to the center-points of each edge. In $H_{cSSPT}$, the global symmetry interchanges the two sums. Note that the edge qubits lie on the vertices of the medial square lattice. 

$H_{cSSPT}$ is exactly a 2D cluster Hamiltonian in an external field, tuned to its critical point~\cite{Doherty2009,Kalis2012}. This critical point corresponds to a first order phase transition between the 2D SSPT phase of the cluster state and the trivial phase~\cite{Stephen2019}. Being first order, the phase transition is caused by a level crossing, such that $H_{cSSPT}$ has two degenerate ground states with a gap above them. In fact, these two grounds states are related by the boundary action of the global symmetry. $H_{SSB}$, on the other hand, corresponds to two decoupled plaquette Ising models~\cite{You2018}, one on the vertical edges, one on the horizontal edges. The ground states of this model spontaneously break the subsystem symmetries, such that there is an extensive number of degenerate ground states, with a gap above them.

We see that both of the above Hamiltonians have a degenerate ground space with a finite gap above. It is interesting to compare this to the analogous situation which arises on the boundary of a 2D SPT order \cite{Levin2012}. In that case, the boundary system is a 1D chain. The minimal Hamiltonian terms that commute with the boundary symmetry correspond to the 1D cluster Hamiltonian in a magnetic field tuned to its critical point, and two decoupled 1D Ising models. Thus the situation is similar to the current one. A crucial difference, however, lies in the fact that the ground state of the critical 1D cluster Hamiltonian respects the anomalous symmetry, and is therefore gapless~\cite{Levin2012}. Furthermore, adding the Ising interaction on top preserves the criticality in a finite region~\cite{Bridgeman2017}. Therefore, the 2D SPT supports a boundary with symmetry-protected gaplessness. Conversely, it is not immediately clear how to engineer a gapless boundary for $|SSPT\rangle$. This is similar to the situation for 2D SSPT phases, which also only support degenerate boundaries~\cite{You2018}. We leave a more detailed analysis of the boundary of 3D SSPT phases to future work.

\section{Calculation of topological entanglement entropy} \label{app:tee}

\begin{figure}[t]
\centering
\includegraphics[width=\linewidth]{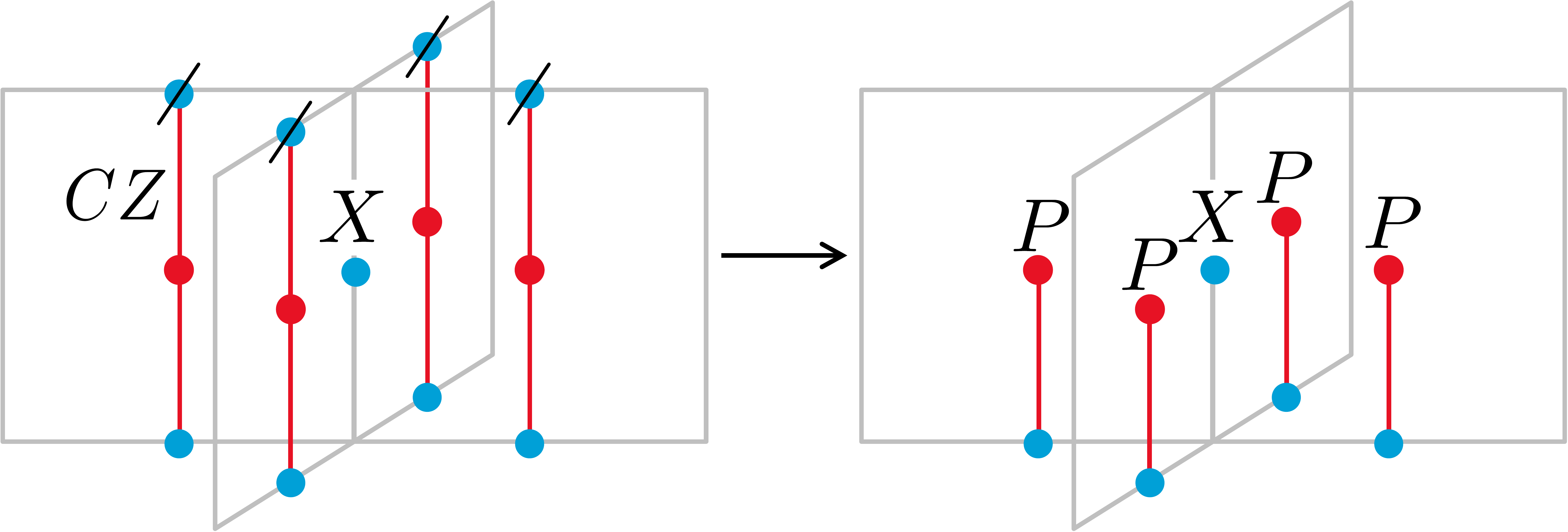}
\caption{After tracing out qubits in the $B$ subsystem, as, indicated by the slashes, the operator $C_e$ (left) is transformed to $C'_eP_e$ up to a constant factor of $2^4=16$.}
\label{fig:traceCZ}
\end{figure}

In this Appendix, we compute $S^{(2)}_A$ for the bipartition depicted in Fig.~\ref{fig:3torus}. The subsystem $A$ is defined by two planar cuts that run parallel to subsystem symmetry planes and intersect edges/faces of the lattice, such that the qubits on intersected edges and faces lie within subsystem $A$ (\textit{ie.} $A$ has rough boundaries rather than smooth ones). The boundary between the two subsystems is therefore two disconnected tori. Let $\partial A_{E}\subset E$ denote the edges intersected by the bipartitioning planes on each end of $A$. Likewise, let $\partial A_{F}\subset F$ denote the intersected faces, and let $\partial A=\partial A_E \cup \partial A_F$. Since the boundary is made of two disconnected pieces, we can break each set into left and right parts, \textit{i.e.}  $\partial A_E=\partial A_E^L\cup \partial A_E^R$ and $\partial A_F=\partial A_F^L\cup \partial A_F^R$.

Let $G$ be the abelian stabilizer group defined in the main text. Then, since $|SSET\rangle$ is the unique state satisfying $g|\psi\rangle=|\psi\rangle$ $\forall g\in G$, we can write \cite{Hamma2005, Zou2016},
\begin{equation}
\rho=|\psi^{3D}_{SSET}\rangle\langle\psi^{3D}_{SSET}|= \frac{1}{2^{|A|+|B|}}\sum_{g\in G} g .
\end{equation}
Where $|A|$ and $|B|$ are the number of qubits in subsystems $A$ and $B$. When we trace out the $B$ subsystem, all elements in $G$ which have non-trivial support in $B$ have zero partial trace, since the Pauli operators are traceless. The only exception are the operators $C_e$ where $e\in\partial A_{E}$. In this case, the trace over $B$ traces over one of the qubits involved in half of the $CZ$ operators in $C_e$. This does not give 0, rather we have $\mathrm{Tr}_a CZ_{ab}=2P_b$ where $P=\frac{\mathbb{1}+Z}{2}$. Using this, we find,
\begin{equation}
\rho_A=\frac{1}{2^{|A|}} \sum_{g\in G_A'} g\sum_{\mathcal{E}\subset \partial A_E} C'_\mathcal{E} P_\mathcal{E} 
\end{equation}
where we have defined the projector $P_\mathcal{E}$ as,
\begin{equation}
P_\mathcal{E}=\prod_{f\in d\mathcal{E}} P_f
\end{equation}
with,
\begin{equation}
C'_\mathcal{E}=\prod_{e\in \mathcal{E}} C'_{e}
\end{equation}
where $C'_{e}$ is a unitary operator obtained from $C_e$ by removing half of the $CZ$'s, see Fig.~\ref{fig:traceCZ}. The subgroup $G_A'$ is defined to be generated by all elements of $G$ which act non-trivially only on $A$, except for the two operators $\prod_{e\in \partial A_E^{L/R}} C_e$, which we exclude for notational convenience. 

To compute $\rho_A^2$, we observe that $[C'_\mathcal{E},g]=[P_\mathcal{E},g]=0$ for all $g\in G_A'$ and $\mathcal{E}\subset \partial A_E$. Then we get,
\begin{equation}
\rho_A^2=\frac{1}{2^{2|A|}}|G_A'|\sum_{g\in G_A'} g \sum_{\mathcal{E},\mathcal{E}'\subset \partial A_E} C'_{\mathcal{E}\oplus\mathcal{E}'} P_\mathcal{E} P_{\mathcal{E}'}
\end{equation}
where we have used the facts $( \sum_{g\in G_A'} g)^2=|G_A'|( \sum_{g\in G_A'} g)$ and 
$C'_\mathcal{E}C'_{\mathcal{E}'}=C'_{\mathcal{E}\oplus\mathcal{E}'}$. We observe that $\rho_A^2$ is not proportional to $\rho_A$ because of the presence of the projectors $P_\mathcal{E}$. Therefore, $\rho_A$ is not a projector, which shows that the entanglement spectrum of our model is not flat, as it would be for the 3D toric code.

Now we take the trace of $\rho_A^2$ in three steps, $\mathrm{Tr}_A=\mathrm{Tr}_{\partial A_F}\mathrm{Tr}_{\partial A_E} \mathrm{Tr}_{A-\partial A}$. When taking the first trace, all non-trivial elements of $G_A'$ with support outside of $\partial A$ are traceless, except again for $C'_e$ for $e\in \partial A_E$. This gives,
\begin{equation}
\mathrm{Tr}_{A-\partial A}(\rho_A^2)=\frac{2^{-|\partial A|}}{2^{|A|}}|G_A'| \sum_{g\in G_{\partial A}'} g \sum_{\mathcal{E},\mathcal{E}'\subset \partial A_E} X_{\mathcal{E}\oplus\mathcal{E}'} P_\mathcal{E} P_{\mathcal{E}'}
\end{equation}
where,
\begin{equation}
X_\mathcal{E}=\prod_{e\in \mathcal{E}} X_e
\end{equation}
is obtained from $C'_\mathcal{E}$ after removing all $CZ$'s. $G_{\partial A}'$ contains all elements of $G_A'$ which have act non-trivially only on $\partial A$, and is generated by the operators $A_e$ for $e\in \partial A_E$. 
Now, the trace over $\partial A_E$ is 0 unless $\mathcal{E}=\mathcal{E}'$, giving,
\begin{equation}
\mathrm{Tr}_{\partial A_E}\mathrm{Tr}_{A-\partial A}(\rho_A^2)=\frac{2^{-|\partial A_F|}}{2^{|A|}}|G_A'|\sum_{g\in G_{\partial A}'} g \sum_{\mathcal{E}\subset \partial A_E} P_\mathcal{E}
\end{equation}
where we used that $P_\mathcal{E} P_\mathcal{E}= P_\mathcal{E}$. At this point, it is useful to consider the two boundaries of $A$ separately. We can write,
\begin{align}
&\mathrm{Tr}_{\partial A_E}\mathrm{Tr}_{A-\partial A}(\rho_A^2)=\frac{2^{-|\partial A_F|}}{2^{|A|}}|G_A'|\,\cdot \nonumber \\
& \sum_{g\in G_{\partial A^L}'} g \sum_{\mathcal{E}\subset \partial A^L_E} P_\mathcal{E}
\sum_{g'\in G_{\partial A^R}'} g' \sum_{\mathcal{E}'\subset \partial A^R_E} P_{\mathcal{E}'} .
\end{align}
Observing that the final trace over $\partial A_F$ can further be factorized as $\mathrm{Tr}_{\partial A_F}=\mathrm{Tr}_{\partial A^L_F}\mathrm{Tr}_{\partial A^R_F}$, and the left and right boundaries are disjoint and equivalent so we can simply square the result for the left boundary, we get,
\begin{align}
\mathrm{Tr}_A(\rho_A^2)&=\frac{2^{-|\partial A_F|}}{2^{|A|}}|G_A'|\, \cdot \nonumber\\
&\left(\sum_{\mathcal{E}\subset \partial A^L_E} \mathrm{Tr}_{\partial A^L_F} \left[P_\mathcal{E}\left( \sum_{g\in G_{\partial A^L}'} g\right)\right]
\right)^2.
\end{align}
For each $\mathcal{E}$ this remaining trace is 0 unless $g=e$, or $g=\prod_{e\in\mathcal{E}} A_e$. In the latter case, we have $P_\mathcal{E}\prod_{e\in\mathcal{E}} A_e=P_\mathcal{E}$. Therefore, every term in the sum over $\mathcal{E}$ gets doubled, except when $\mathcal{E}=\emptyset$ or $\mathcal{E}=\partial A^L_E$. This gives,
\begin{align}
\mathrm{Tr}_A(\rho_A^2)&=\frac{2^{-|\partial A_F|}}{2^{|A|}}|G_A'| \nonumber \\ &\left(2\left[\sum_{\mathcal{E}\subset \partial A_E^L} \mathrm{Tr}\,P_\mathcal{E}\right] - \mathrm{Tr}\, P_\emptyset - \mathrm{Tr}\, P_{\partial A^L_E}
\right)^2 .
\end{align}
Noting that $\mathrm{Tr}\, P_\emptyset = \mathrm{Tr}\, P_{\partial A^L_E} = 2^{|\partial A^L_F|}$, we write,
\begin{align} \label{eq:ee_1}
\mathrm{Tr}_A(\rho_A^2)&=\frac{4}{2^{|A|}}|G_A'| \nonumber \\ & \left(2^{-|\partial A_F^L|} \sum_{\mathcal{E}\subset \partial A_E^L} \mathrm{Tr}\,P_\mathcal{E} - 1
\right)^2 .
\end{align}

We now proceed by expressing the remaining sum in terms of the partition function of a 2D square lattice Ising model. To this end, we define an auxillary square lattice system with degrees of freedom $\sigma_i=\pm 1$ on each vertex near the boundary. Each vertex $i$ corresponds to an edge $e_i \in\partial A_E^L$. Given $\mathcal{E}$, we define a vector $\vec{\sigma}$ such that $\sigma_i=-1$ if $e_i\in \mathcal{E}$, and $\sigma_i=1$ otherwise. With this notation, we can rewrite,
\begin{equation}
P_\mathcal{E}=\prod_{\langle i,j \rangle} \left(P_{l(e_i,e_j)}\right)^\frac{1-\sigma_i\sigma_j}{2}
\end{equation}
where $l(e_i,e_j)$ refers to the face that links edges $e_i$ and $e_j$. The trace is given by,
\begin{equation}
\mathrm{Tr}\, P_\mathcal{E}=\prod_{\langle i,j \rangle} 2\left(\frac{1}{2} \right)^\frac{1-\sigma_i\sigma_j}{2}=\sqrt{2}^{|\partial A^L_F|} \prod_{\langle i,j \rangle} \sqrt{2}^{\sigma_i\sigma_j}
\end{equation}
such that,
\begin{equation}
\sum_{\mathcal{E}\subset \partial A_E^L} \mathrm{Tr}\,P_\mathcal{E}=\sqrt{2}^{|\partial A^L_F|}\sum_{\vec{\sigma}} e^{\ln \sqrt{2} \sum_{\langle i,j\rangle} \sigma_i\sigma_j} .
\end{equation}
We have written the sum in such a way that it corresponds to the partition function $\mathcal{Z}(\ln \sqrt{2})$ of the 2D Ising model, where,
\begin{equation}
\mathcal{Z}(\beta)=\sum_{\vec{\sigma}} e^{\beta \sum_{\langle i,j\rangle} \sigma_i\sigma_j} .
\end{equation}
With this, Eq.~\ref{eq:ee_1} becomes,
\begin{equation}
\mathrm{Tr}_A(\rho_A^2)=\frac{4}{2^{|A|}}|G_A'| \left(\sqrt{2}^{-|\partial A_F^L|} \mathcal{Z}(\ln\sqrt{2}) - 1
\right)^2 .
\end{equation}

For conceptual clarity, let us define the group $G_A$ which is obtained by adding the two generators $\prod_{e\in \partial A_E^{L/R}} C_e$, which we omitted earlier, into $G_A'$. Then we have $|G_A|=4|G'_A|$. By letting $N=|\partial A_E^L|$ be the number of spins in the Ising model, we get,
\begin{align}
S^{(2)}_A&=-\ln \mathrm{Tr}_A(\rho_A^2) \nonumber \\
&=|A|\ln 2-\ln |G_A| -2\ln(2^{-N} \mathcal{Z}(\ln \sqrt 2)-1) .
\end{align}
This Ising model has a phase transition at inverse temperature $\beta_c=\frac{\ln (1+\sqrt{2})}{2}$ \cite{Onsager1944}. Since $\ln\sqrt{2}<\beta_c$, our partition function lies in the disordered phase. In this phase, the free energy is extensive, in the sense that $\mathcal{F}(\beta):=\ln Z(\beta)=\alpha N$ for large $N$. Importantly, there is no constant term in $\ln Z(\beta)$, as there would be in the ordered phase. The 2D Ising model has been solved exactly in the large-$N$ limit by Onsager \cite{Onsager1944}, and is given in Ref.~\cite{Somendraw95} as,
\begin{align}
\alpha =& \ln\left(2\cosh(2\beta)\right)\  + \nonumber \\
&\frac{1}{2\pi}\int_0^\pi d\phi\  \ln\frac{1}{2}\left(1+\sqrt{1-k^2 \sin^2\phi} \right)
\end{align}
where $k = 2 \sinh(2\beta)/\cosh^2(2\beta)$. Evaluating this integral numerically for $\beta = \ln\sqrt{2}$, we find that $\alpha$ is equal to $\ln 2(\ln 2 + \frac{1}{2})$ up to 6 decimal places.
Thus for large values of $N$ we get
\begin{align}
S^{(2)}_A&=-\ln \mathrm{Tr}_A(\rho_A^2) \nonumber \\
&=|A|\ln 2-\ln|G_A|-2\ln\left( e^{\ln 2(\ln 2 - \frac{1}{2})N}-1\right) .
\end{align}
Note that $e^{\ln 2(\ln 2 - \frac{1}{2})}>1$, such that we can approximate $\ln\left( e^{\ln 2(\ln 2 - \frac{1}{2})N}-1\right)\approx\ln\left( e^{\ln 2(\ln 2 - \frac{1}{2})N}\right)$ for large $N$, giving,
\begin{equation}
S^{(2)}_A=|A|\ln 2-\ln|G_A|-2\ln 2(\ln 2 - \frac{1}{2})N\ .
\end{equation}

The final step is to evaluate $|G_A|$. To do this, we need to count the number of indepedent generators of $G_A$. This is done using the usual counting arguments for the 3D toric code \cite{Castelnovo2008}.  Suppose for simplicity that the region $A$ contains $L\times L\times L$ vertices, such that $N=L^2$. We then have $3L^3+L^2$ edges contained in $A$, and $3L^3+2L^2$ faces, giving $|A|=6L^3+3L^2$. Starting with the body center terms $B_c$, we have $L^3-L^2$ terms contained in $A$, all of which are independent. We have $3L^3$ edge terms $A_e$, but they are not all independent. Namely, the product of all edge terms around a given vertex is the identity, and so is the product of all edge terms on a non-contractible plane (such as the plane $\Sigma_z$ in Fig.~\ref{fig:3torus}, giving $L^3+1$ constraints, and hence $2L^3-1$ independent terms. Finally, we have one $C_e$ term for each edge away from the boundary, all of which are independent, and the two non-local terms $\prod_{e\in \partial A_E^{L/R}} C_e$, giving $3L^3-L^2+2$ terms. Finally, we have the three non-local operators  ${S}^\mathrm{m}_{\Sigma_{z}}$ and $S^{\mathrm{e}}_{\Lambda_{x,y}}$. Altogether, this gives $6L^3-2L^2+4$ terms, such that $|G_A|=2^{6L^3-2L^2+4}$. 
The final result for the entropy is therefore,
\begin{equation}
S^{(2)}_A(L)=\ln 2(6-2\ln 2) L^2-4\ln 2\ ,
\end{equation}
plus corrections which go to zero as $L$ goes to infinity.

\end{document}